\documentclass[
article,
onecolumn,
]{IEEEtran}

\usepackage[T1]{fontenc}%

\usepackage[english]{babel}	%
\usepackage{amsmath,amsfonts}					%
\usepackage{graphicx}									%
\usepackage[svgnames]{xcolor}							%
\usepackage{epstopdf}												%
\usepackage{booktabs}												%
\usepackage[square,sort,comma,numbers,compress]{natbib}
\usepackage{hyperref}
\usepackage{url}

\usepackage[ruled,vlined,commentsnumbered,linesnumbered,lined,boxed]{algorithm2e}
\usepackage{algorithmicx}
\usepackage{algpseudocode}

\usepackage[hang, small, labelfont=bf, textfont=it, up]{caption}	%
\usepackage[skip=0cm,list=true]{subcaption}
\usepackage{paralist}
\usepackage{flushend}
\usepackage{balance}
\usepackage{psfrag}

\SetKwProg{Fn}{Function}{}{}

\begin{document}

\title{T-RACKs: A Faster Recovery Mechanism for TCP in Data Center Networks}

\newcommand*{\affaddr}[1]{#1}
\newcommand*{\affmark}[1][*]{\textsuperscript{#1}}
\newcommand*{\email}[1]{\texttt{#1}}

\author{
Ahmed M. Abdelmoniem\affmark[1,2] \quad Brahim Bensaou\affmark[2]
\\
\affaddr{\affmark[1]CS Department, Assiut University, Egypt}\\
\affaddr{\affmark[2]CSE Department, HKUST, Hong Kong}\\
\email{\{amas,brahim\}@cse.ust.hk}
\thanks{Manuscript is accepted for publication in ACM/IEEE Transactions of Networking~\copyright{}2021 IEEE }
}

\maketitle

\begin{abstract}
Cloud interactive data-driven applications generate swarms of small TCP flows that compete for the small buffer space in data-center switches. Such applications require a short flow completion time (FCT) to perform their jobs effectively. However, TCP is oblivious to the composite nature of application data and artificially inflates the FCT of such flows by several orders of magnitude. This is due to TCP's Internet-centric design that fixes the retransmission timeout (RTO) to be at least hundreds of milliseconds. To better understand this problem, in this paper, we use empirical measurements in a small testbed to study, at a microscopic level, the effects of various types of packet losses on TCP's performance. In particular, we single out packet losses that impact the tail end of small flows, as well as bursty losses, that span a significant fraction of the small congestion window of TCP flows in data-centers, to show a non-negligible effect on the FCT. Based on this, we propose the so-called, timely-retransmitted ACKs (or T-RACKs), a simple loss recovery mechanism to conceal the drawbacks of the long  RTO even in the presence of heavy packet losses. Interestingly enough, T-RACKS achieves this transparently to TCP itself as it does not require any change to TCP in the tenant's virtual machine (VM). T-RACKs can be implemented as a software shim layer in the hypervisor between the VMs and server's NIC or in hardware as a networking function in a SmartNIC. Simulation and real testbed results show that T-RACKs achieves remarkable performance improvements.
\end{abstract}

\begin{IEEEkeywords}
Data-center, Cross Layer, Fast Recovery, Kernel Module, TCP-Incast, Timeouts.
\end{IEEEkeywords}

\IEEEpeerreviewmaketitle

\newcommand{\SWITCH}[1]{\STATE \textbf{switch} (#1)}
\newcommand{\ENDSWITCH}{\STATE \textbf{end switch}}
\newcommand{\CASE}[1]{\STATE \textbf{case} #1\textbf{:} \begin{ALC@g}}
\newcommand{\ENDCASE}{\end{ALC@g}}
\newcommand{\CASELINE}[1]{\STATE \textbf{case} #1\textbf{:} }
\newcommand{\DEFAULT}{\STATE \textbf{default:} \begin{ALC@g}}
\newcommand{\ENDDEFAULT}{\end{ALC@g}}
\newcommand{\DEFAULTLINE}[1]{\STATE \textbf{default:} }

\section{Introduction}
\label{sec:introduction}

The recent growth in data-center deployments worldwide is reshaping how the Internet and its applications operate. New, cloud-based, data-driven applications have emerged over the past many years to harness the cost-effectiveness and scalability afforded by cloud computing. Most such applications rely on distributed programming and data storage frameworks such as Hadoop, HDFS, or Spark \cite{Zaharia2010} for storing and processing large data sets. In such frameworks, master and aggregation nodes often require data transfers from hundreds of worker nodes to build a result. Due to the stringent timing requirements of interactive applications, a data transfer that misses a hard deadline because of excessive waiting for packet loss recovery returns a \textbf{partial} result (of lower quality). Hence, the quality of application results is not only correlated with the average latency but also with the tail of latency distribution. For example, in practice, the $90^{th}$\% of the flow completion times (FCT) can be anywhere between two to four orders of magnitude worse than the median or the average latency. 

In small scale private data-centers, CPU resources are often the bottleneck, and solutions that rely on task placement and scheduling already exist (e.g., \cite{Mattess2013}). In contrast, public data-centers seldom overload their server CPUs and usually have abundant computing resources; yet, they often adopt high over-subscription ratios in the network. As a result, network latency becomes the main performance bottleneck \cite{Rumble2011}. This is typical for many Internet-scale applications deployed on public (IaaS) clouds such as Microsoft Azure or Amazon EC2.

Measurements in real production data-centers \cite{Phanishayee2008, Kandula2009, Alizadeh2010, Wu2013, Judd2015} have shown over the years that the applications that produce small traffic flows predominate and that incast congestion events and excessive packet losses are frequent. To circumvent such problems, large corporations such as Microsoft, Facebook, and Google dedicate well-structured data-centers to deploying their time-sensitive applications. Smaller-scale private data-centers address the problem by deploying homogeneous custom-designed TCP variants (e.g., DCTCP \cite{Alizadeh2010}) on all the VMs in the data-center. In stark contrast, multi-tenant and public data-centers, where many-to-one (or many-to-many) communication patterns predominate, are populated with a large variety of versions of TCP with different behaviors in the face of congestion~\cite{Kandula2009, Greenberg2009, Benson2010}. As a direct consequence of this heterogeneous environment, unfairness in congestion resolution is inevitable and often leads to repeated packet losses and a long-tailed latency distribution for small flows. In particular, since commodity Ethernet switches are the backbone of all intra-data-center communications, their small buffer space can quickly be fully occupied by a few (large) TCP flows. This is because TCP has a natural tendency to fill up the bottleneck bandwidth of the communication path.  

This raises two issues: 
\begin{inparaenum}[\itshape i)\upshape] 
\item small flows would not last long enough to capture their fair share of the buffer from ongoing large flows, as their TCP sending window cannot grow large enough before a packet loss is experienced; 
\item when a sudden swarm of small flows (usually co-flows) surges, while the buffer is occupied by other flows, incast congestion loss events become inevitable. In this case, a burst of packet losses from many such flows takes place. Bursty losses with small congestion windows often leave an insufficient number of TCP segments in flight to trigger TCP's fast-retransmit and recovery mechanism. As a consequence, small flows often experience timeouts. The retransmission timeout (RTO), which is orders of magnitude larger than the actual round-trip time (RTT), contributes thus the lion's share to the long FCT and number of missed deadlines experienced by small flows in data-centers. 
\end{inparaenum}

In this paper, we study the impact of RTO on the performance of TCP applications in data-centers and propose a simple mechanism to shield small TCP flows from the negative effects of the RTO, without changing TCP itself. Our methodology adopts a two-phased approach:

\begin{inparaenum}[\itshape i)\upshape]
\item First, to fully understand the impact of the RTO on the FCT of small flows, we conduct an empirical study of the loss events in a small data-center, by examining the nature of the recovery mechanism invoked by TCP for each segment loss.  To this end, we trace TCP traffic flows microscopically at the socket-level in the Linux Kernel. Then by analyzing the collected traces, we study the frequency of occurrence of the two TCP loss recovery mechanisms (viz., RTO\footnote{The minimum RTO is 200ms in Linux and 300ms in Windows} and Fast Retransmit and Recovery (or FRR)) concerning the TCP window size. We show that tail-end losses and bursty losses primarily cause RTOs, and while they have a less dramatic effect on the latency of large flows, their impact on the performance of small flows is tremendous. 
\item Second, to prevent RTOs from artificially inflating the actual loss recovery delays of small flows intra-data-centers, and without modifying TCP\footnote{Notice that in public data-centers, under the IaaS model, the operating system and thus the protocol stack in the VM is under the full control of the tenant and cannot be modified by the cloud service provider.} we propose, implement and study the performance of a new mechanism to conceal the long retransmission timeout. This mechanism forces TCP in the VM to go into the FRR mode whenever a segment is estimated to be likely to experience a timeout, long before it does. We implement the resulting so-called T-RACKS in a real testbed and study its performance with realistic traces\footnote{An earlier version of this work were published in INFOCOM 2018~\cite{TRACKS-INFOCOM}. \\The implementation, simulation and experimental code and scripts are publicly available at \url{http://ahmedcs.github.io/T-RACKs}.}.
\end{inparaenum}

In the remainder, supported by an empirical study, we show in Section \ref{sec:problem} the dramatic impact of the RTO on the performance of small flows. In Section \ref{sec:system}, we present the proposed methodology and system design. In Section \ref{sec:sim}, we discuss the packet-level simulation results in detail. Then, in Section \ref{sec:imp}, we present the experimental results from the testbed deployment. We discuss important related work in Section \ref{sec:related}. Finally, we conclude the paper in Section \ref{sec:conclude}. %

\section{Related Work}
\label{sec:related}
Several works have found, via measurements and analysis, that TCP timeouts are the root cause of most throughput and latency problems in data-center networks \cite{Vasudevan2009, Shukla2014,Ahmed-IPCCC-2015,Ahmed-ANNALS-2017,Ahmed-ITCE-2019,Ahmed-INFOCOM-2019,Ahmed-ICDCS-2019-1,Ahmed-ICPP-2020}. Other works \cite{Chen2009,Zhang2011,Tam2012, Zats2012, JiaoZhang2013, Irteza2014, Chen2015, Zhang2015,Ahmed-ICC-2016-1,Ahmed-LCN-2017,Ahmed-SRDS-2017,Ahmed-GLOBECOM-2018} analyzed the nature of incast events and packet drops in data-centers\footnote{Similar analysis was done for Content Centric Networks (CCNs)~\cite{Ahmed-ICC-2016-2,Ahmed-LCN-2016}},. They also found that severe incast occurrences could lead to throughput collapse and longer FCT. They show in particular that throughput collapse and increased FCT are to be attributed to the data-center ill-suited timeout mechanism. For example, \cite{Vasudevan2009} showed that frequent timeouts could harm the performance of latency-sensitive applications. Numerous solutions have been proposed. These fall into one of four fundamental categories. The first mitigate the consequence of the long waiting times due to RTO, by reducing the default $RTO_{min}$ to the 100 $\mu$s - 2 ms \cite{Vasudevan2009}. While very useful, this approach affects the sending rates of TCP by forcing it to cut CWND to 1; it relies on a static $RTO_{min}$ value, which can be ineffective in heterogeneous networks; and it imposes modifications to TCP stack on tenant's VM. Our approach is fundamentally different in its enforcement of $RTO_{min}$ via dup-acks which allows for different handling of Internet and data-centers flows. Therefore, T-RACKs allows flows to have different $RTO_{min}$ which is easily imposed by the flow tables.

The second approach aims at controlling queue build-up at the switches by relying on ECN marks to limit the sending rate of the servers \cite{Alizadeh2010,Judd2015,Ahmed-LCN-2016}, or by controlling the congestion window \cite{JiaoZhang2013} or receiver window \cite{Wu2013,Ahmed-CLOUDNET-2015,Ahmed-GLOBECOM-2015} of TCP flows. Similar approaches deployed global traffic scheduler \cite{Alizadeh2012,Alizadeh2012a,Bai2015,Ahmed-ICC-2019,Ahmed-ICDCS-2019-2} or tacked fine-grained sub-microsecond updates in RTT to detect congestion \cite{Mittal2015}. All of these works achieved their goals and have shown they could reduce the FCT of short flows as well as achieving high link utilization. However, they require modifications of either the TCP stack, or introduce a completely new switch design, and are prone to fine-tuning of parameters or sometimes require application-side information. They also increase CPU utilization of the end hosts. \cite{Mittal2015} is sensitive to traffic variations in the backward path.~\cite{Cardwell2017} is a new congestion control for inter-DC traffic which is based on characterizing the band-width and RTT of the bottleneck path. While effective for high bandwidth-delay product (BDP), its minutes-level measurements and the aggressive start can exacerbate the problems in low BDP networks of data-centers.

The third approach is to achieve efficient sharing of network resources or enforce flow admission control to reduce TimeOut probability~\cite{Benson2011,Popa2012,Shukla2014,Huang2016}. \cite{Huang2016} has proposed ARS, a cross-layer system that can dynamically adjust the number of active TCP flows by batching application requests based on the sensed congestion state indicated by the transport layer. The last approach, which is adopted in this paper due to its simplicity, and feasibility, is to recover losses utilizing fast retransmit rather than waiting for a long timeout. For instance, TCP-PLATO \cite{Shukla2014} proposed changing TCP state-machine to tag specific packets using IP-DSCP bits, which are preferentially queued at the switch to reduce their drop-probability; enabling dupACKs to be received to trigger FRR instead of waiting for the timeout. Even though TCP-PLATO is effective in reducing timeouts, its performance is degraded whenever tagged packets are lost. In addition, the tagging may interfere with the operations of middle-boxes or other schemes, and most importantly, it modifies the TCP state machine of the sender and receiver. 

Similar to DCTCP, DCQCN \cite{Alizadeh2008,Alizadeh2011} and HPCC~\cite{Yuliang2019} was proposed as an end-to-end congestion control scheme implemented in custom NICs designed for RDMA over Converged Ethernet (RoCE). Both DCQCN, and HPCC applies adaptive rate control at the link-layer to throttle large flows relying on Priority-based Flow Control (PFC) and RED-ECN marking, and In-Network Telemetry (INT) information, respectively. DCQCN, not only relies on PFC, which adds to network overhead, it introduces the extra cost of the explicit ECN Notification Packets between the end-points. HPCC requires programmable NICs and relies on the timely availability of the INT information which not only increases the packet size by 42 bytes for each hop in the path but also is subject to congestion, and contention with other traffic in the network.

More recent approaches have also identified the importance of the timeout problem in data-center environments \cite{Chang2019, Xu2019, Zhaung2019}. For instance, the authors in \cite{Chang2019} proposed injects high-priority packet after each window worth of packets. They infer the network congestion by checking the sequences of the received ACKs of high/low priorities. Consequently, they adjust the sending window to reduce buffer occupancy and early detect losses. However, this not only requires setting-up priority-queues (if available) in the switches but also imposes extra processing and communication overhead (esp., the congestion window consists of typically few packets in data-centers).

\section{Problem and Motivation}
\label{sec:problem}

Before we start discussing our empirical study of TCP and presenting our solution, let us first shed some light on the motives that led us to adopt such a non-traditional approach by contrasting it against alternative methods. In particular, while our approach is straightforward, it turns out to be very effective because it stems from a full understanding of the large number of incremental mechanisms that have been added over the years to TCP. 
Many alternative schemes proposed in the literature deal with TCP congestion in data-centers in a classic Internet-centric approach by invoking mechanisms such as RED. This approach is flawed because of three major reasons:
\begin{inparaenum}[\itshape i)\upshape]
\item RED is a mechanism that was designed for the Internet. Its goal is to reduce the average queuing delay experienced by packets in the huge routers' buffer, which contributes a large proportion of the end-to-end delay and delay-jitter. In contrast, data-centers use high-speed switches with small buffers; therefore, the contribution of queuing delay to the total FCT is not as dramatic as in the case of the Internet, regardless of the buffer occupancy. And so, maintaining a small queue does not help the FCT.
\item With increasing link speeds in modern data-centers, the interplay between propagation delay and transmission delay is transformed, rendering the control mechanisms valid for one no longer valid for the other. For example, in data-centers with 1Gbps network interfaces, the transmission time of a single IP packet of 1500 bytes is about 12 microseconds; the round trip time over a 600m path at the speed of light is about 6 microseconds. So, there can be at most one packet spread over a link between any two adjacent interfaces in the network (e.g., server NIC, to ToR port, ToR port, to Aggregation Port, ...). In contrast, with 40 Gbps network interfaces, it takes only 0.3 microseconds to complete the transmission of an IP packet, yielding a possible flight of up to 33 IP packets per hop. So early detection and notification via buffer thresholds with the small buffers that exist in the switches are ineffective, and excessive packet losses are inevitable. 
\item Packet losses in TCP per-se are not the reason for these problems; excessive congestion is. Packet losses are merely symptoms of congestion, so there is no reason to try to curtail them completely as long as we can recover from losses fast. In fact, curtailing packet losses completely results in a non-competitive behavior that yields poor performance in heterogeneous TCP environments. For example, pitting TCP Vegas against TCP New Reno, Cubic or DCTCP results in poor performance for the former.
\end{inparaenum}
As a consequence, eliminating packet losses in data-center networks while maintaining a high link utilization is not helpful. Instead, we propose to pinpoint the true reason for increased delays in data-centers and to tackle such reasons directly~\cite{TRACKS-INFOCOM}.

Several measurement studies \cite{Kandula2009, Greenberg2009, Benson2010} have been conducted on data-centers and have shown that latency in such environments varies greatly. To further understand the reasons behind this, we deep-dive into the packet level analysis of the flows and the TCP socket state variables at a microscopic level to understand TCP behavior and its loss recovery mechanisms. 
An early work \cite{Vasudevan2009}, based on data-center measurements, found that the timeout mechanism is to blame for the long waiting times and proposed the very simple yet effective solution of reducing the $RTO_{min}$ value for TCP in data-center environments while using high-resolution timers to keep track of delays at the microsecond-level. This approach actually solves the problem, reduces the FCT, and mitigates TCP-incast congestion effects. However, 
\begin{inparaenum}[\itshape i)\upshape]
\item it requires the modification of TCP, and as such it is inappropriate for public data-centers where multiple tenants can upload their own version of the OS; and, 
\item there is no ``magical'' value of $RTO_{min}$ that fits all possible environments. For instance, a $RTO_{min}$ that works inside the data-center (e.g., between a web server and the back-end database server) will lead to spurious timeout events for Internet-facing connections (e.g., the connection between the web administrator workstation and the server in the data-center).
\end{inparaenum}

A recent RFC \cite{TLP} proposed the so-called tail loss probe (TLP) mechanism, which recommends sending TCP probe segments whenever ACKs do not arrive within a short Probe TimeOut (PTO)\footnote{PTO is set to min(2*srtt, 10ms) if inflight$>$1 and to 1.5*srtt + worst case delayed ACK (i.e., 200ms) if inflight==1}. In addition to requiring changes to TCP, this approach suffers from two additional problems: 
\begin{inparaenum}[\itshape i)\upshape]
\item probe packets also may be lost; and, 
\item probe packets may worsen the in-network congestion, especially during TCP-incast.
\end{inparaenum}

\subsection{Impact of RTO on The FCT}
In data-centers, partition/aggregate applications that generate small flows are challenged by the presence of small buffers, large initial sending windows, inadequate $RTO_{min}$, or slow-start exponential increase. This combination of hardware and TCP configuration frequently leads to timeout events for such applications. In particular, when the number of flows they generate is large and roughly synchronized, incast-TCP synchronized losses occur. As the loss probability increases linearly with the number of flows \cite{Mathis1997}, the flow synchronization and the excessive losses lead to throughput-collapse for small-flows. 

To illustrate this, consider a simplified fluid-flow model with $N$ flows sharing a link of capacity $C$ equally. Let $B$ be the flow size in bits and $n$ be the number of RTTs it takes to complete the transfer of one flow. The optimal throughput $\rho^*$ can be simply expressed as the fraction of the flow size to its average transfer time: $\rho* = \frac{B}{n \tau + \frac{B N}{C}}$. That is, it takes $BN/C$ to transmit the $B$ bits plus an additional queuing and propagation delay of $\tau$ seconds for each of the $n$ RTTs. In practice, when TCP incast congestion involving $N$ flows results in throughput-collapse, the flow experiences one or more timeouts and recovers after waiting for RTO. Then, the actual throughput writes: $\rho = \frac{B}{RTO + n^\prime \tau^\prime +\frac{B N}{C}}$, Typically $n^\prime \geq n$ and $\tau^\prime \geq \tau$. In addition, in data-centers, the typical RTT is around $100\mu$s, while existing TCP implementations impose a minimum RTO (i.e., $RTO_{min}$) of about $100$ to $200$ms\footnote{Linux uses 200ms and Windows uses 300ms}. As a consequence, large flows yield values of $n^\prime$ such that $n^\prime\tau^\prime$ is similar or greater than $RTO$. In contrast, small flows only last for a few RTTs, therefore $RTO \gg n^\prime\tau^\prime$. And so, when a small flow experiences a loss that cannot be recovered by 3-duplicate ACKs, it systematically incurs an FCT that is orders of magnitude larger than it should.

\subsection{Analyzing TCP congestion recovery}

\begin{figure}[!t]
\captionsetup[subfigure]{justification=centering}
\centering
              \begin{subfigure}[ht]{0.48\columnwidth}
       \includegraphics[width=\columnwidth]{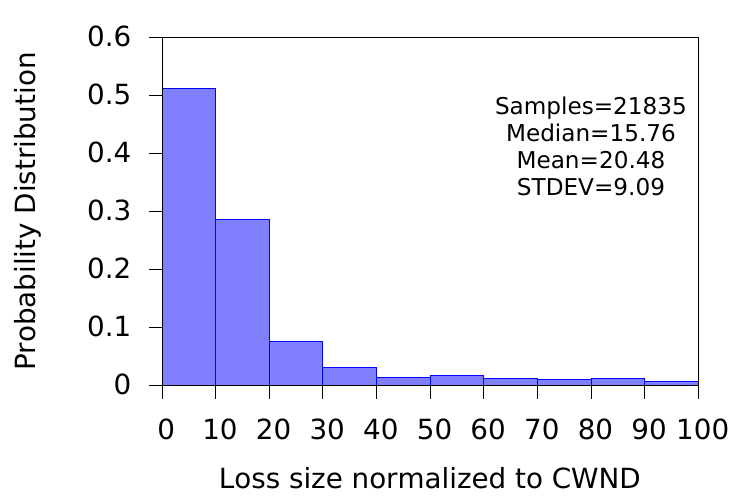}
	\caption{FR size rel. CWND size}
            \label{fig:analysis-frsize}
           \end{subfigure}
    	\hfill
    \begin{subfigure}[ht]{0.48\columnwidth}
       \includegraphics[width=\columnwidth]{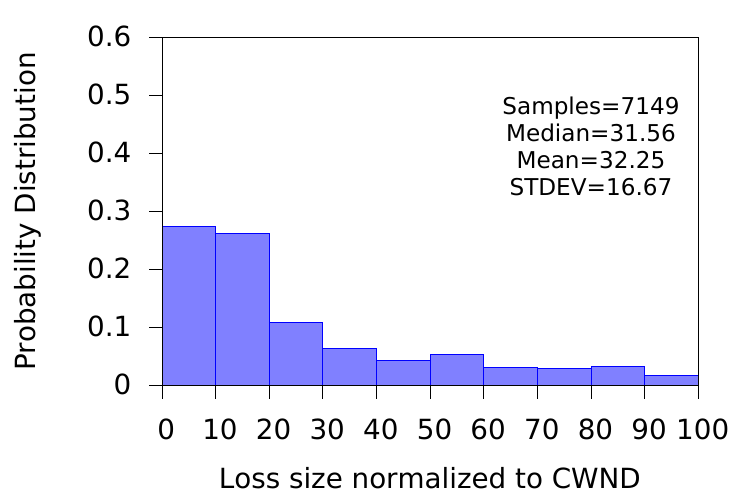}
	\caption{RTO size rel. CWND size}
      \label{fig:analysis-rtosize}
       \end{subfigure}
       \\

    \begin{subfigure}[ht]{0.48\columnwidth}
           \includegraphics[width=\columnwidth]{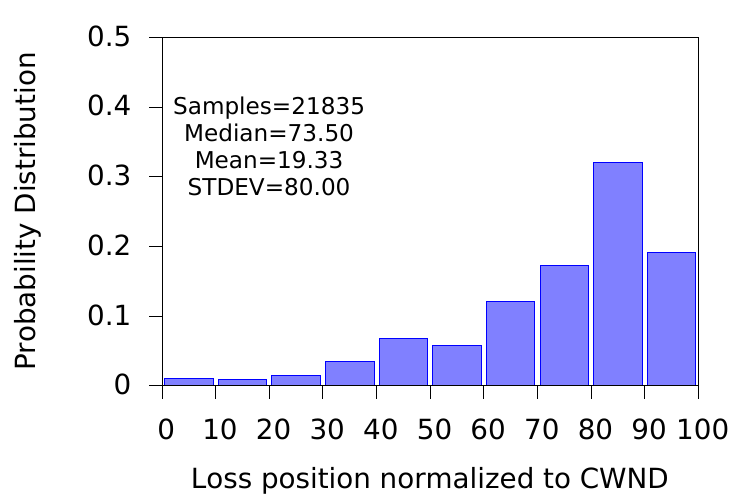}
	\caption{FR position rel. CWND}
	\label{fig:analysis-frpos}
	\end{subfigure}
                \hfill
     \begin{subfigure}[ht]{0.48\columnwidth}
           \includegraphics[width=\columnwidth]{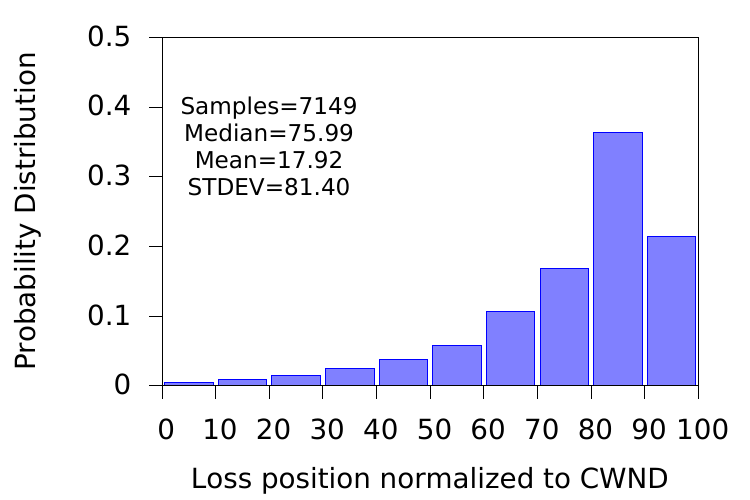}
	\caption{RTO position rel. CWND}
	\label{fig:analysis-rtopos}
       \end{subfigure}

	\caption{(a-b) shows the retransmission size relative to CWND while (c-d) shows the loss position relative to CWND.}
	\label{fig:analysis}

\end{figure}

\begin{table}[!t]
\centering
 \caption{TCP API Calls Intercepted by LossProbe Module}
 \label{tab:apis}
\begin{tabular}{ |p{2.5cm}|p{5cm}| }

 \hline
Function Call & Description\\
 \hline
tcp\_set\_state  &  Handle and update TCP connection state \\
tcp\_v4\_do\_rcv & Handles the arrival of all types of TCP segments \\
tcp\_retransmit\_skb & Retransmits one SKB where policy decisions and retransmit queue state updates are done by the caller\\
tcp\_v4\_send\_check & Computes an IPv4 TCP checksum \\
 \hline
\end{tabular}
\end{table}

\begin{table}[!t]
\centering
 \caption{TCP Socket-level State Info logged by LossProbe Module}
 \label{tab:states}

\begin{tabular}{ |p{1.1cm}|p{2cm}|p{4.5cm}| }
 \hline
Data Type & Variable & Description\\
 \hline
 uint & lost\_out & Count of lost packets\\
 uint & prr\_out & Count of pkts sent during Recovery\\
 uint & prr\_delivered & Count of packets delivered during recovery\\
 uint & prior\_cwnd & Congestion window at start of recovery\\
 uint & prior\_ssthresh & SSThresh saved at recovery start\\
 uint & total\_retrans & Count of retransmits for entire connection\\
 uint & retransmit\_high & Highest sequence \# of retransmitted data\\
 uint & lost\_retrans\_low & Lowes sequence \# of retransmitted data\\
 uint & packets\_out & Count of segments currently in flight\\
 uint & retrans\_out & Count of packets retransmitted\\

  int               & undo\_retrans & Count of undo-able retransmissions\\
  int               & rcv\_tstamp &  Timestamp of last received ACK (for keep-alive)\\
  int               & lsndtime & Timestamp of last sent data packet (for restart window)\\
  int               & retrans\_stamp & Timestamp of the last retransmit\\
  int               & recovperiod & Duration of total recovery period\\
  int               & maxrecovperiod &  Maximal recovery time experienced \\
 \hline
\end{tabular}
\end{table}

To investigate why packet losses seem to affect large flows only marginally, yet degrade the performance of small flows dramatically, we collected and examined socket-level TCP flows state information from a Websearch workload~\cite{Alizadeh2010}, in a small-scale data-center testbed. First, we implemented a socket-level monitoring module, named hereafter ``LossProbe'', based on KProbes/JProbes~\cite{jprobe} in the Linux Kernel. Probes are dynamic debugging tools which in our case, allow us to intercept different TCP event handlers and API calls as listed in Table~\ref{tab:apis} where we log the target TCP socket-level state information. The module works as follows:
\begin{enumerate}
\item Jprobe requires the address of the kernel function to trace; hence the target TCP handlers of the events of interest have to be identified from the Linux kernel source code base~\cite{linuxcodebase}. For example, \textbf{tcp\_retransmit\_skb} is the function called  in the kernel to retransmit a TCP segment.
\item Then, a handler function is defined that will perform certain actions upon entry of the traced function (e.g., print the debugging message when the target kernel function is invoked). In the probe module, that function is defined for convenience with the same name as the original probed function (e.g., \textbf{jtcp\_retransmit\_skb}) and jprobe calls it upon entering the original function.
\item The monitoring module is dynamically installed into the kernel, and the probed workload (or experiment) is invoked. The module upon entry of the targeted functions, the jprobe function, is called to collect the state information of interest and write them into an in-RAM buffer which in turn is flushed periodically and asynchronously onto the file system to avoid stalling the  datapath artificially.
\end{enumerate}
Using our custom-built traffic generator, we replicate a Websearch workload~\cite{Alizadeh2010} consisting of thousands of flows and collect measurements on the data listed in Table~\ref{tab:states} from all the servers in our testbed\footnote{Code for the LossProbe module is publicly available at \url{https://github.com/ahmedcs/TCP_loss_monitor/}}. 

We summarize our findings in several figures to reflect TCP's behavior with respect to the mechanism invoked to recover from packet losses (e.g. Fast Retransmit and Recovery or Re-transmission Timeouts\footnote{Each figure shows the aggregate of all servers in the data-center.}). Fig.~\ref{fig:analysis-frsize} shows the distribution (on the ordinate) of the size of each retransmission (on the abscissa) for the FRR-based recovery events. The size of retransmission is calculated by subtracting the seq\# of the first segment from that of the last segment in a single recovery round which is then normalized by the size of the congestion window $Cwnd$\footnote{The bar for 0-10, refers to probability of 0-10\% of CWND is lost while a bar for 90-100 refers to the probability that 90-100\% of the CWND is lost}. Similarly, Fig.~\ref{fig:analysis-rtosize} shows the same metric for RTO-based recovery events. Fig.~\ref{fig:analysis-frpos}~and Fig.~\ref{fig:analysis-rtopos} show the distribution (on the ordinate) of the loss index or position (on the abscissa) in case of FRR and RTO-based recovery, respectively. The index points to the first retransmitted segment (for several consecutive segments losses), relative to $Cwnd$ when the segment was first transmitted (i.e., before a loss is detected)\footnote{A bar for 0-10, refers to the probability that loss occurs for the first 0-10\% of the segments in $Cwnd$. In contrast, a bar for 90-100 refers to the probability that the loss occurred for the last 90-100\% of the segments the $Cwnd$}. 

\begin{figure}[!t]
\captionsetup[subfigure]{justification=centering}
\centering
 \begin{subfigure}[ht]{0.48\columnwidth}
    \includegraphics[width=\textwidth]{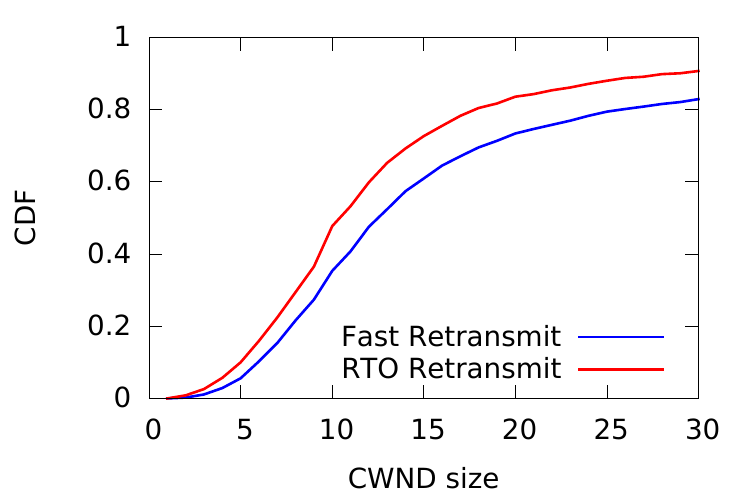}
	\caption{CDF of CWND size}
	\label{fig:cwndsize}
       \end{subfigure}
       \hfill
    \begin{subfigure}[ht]{0.48\columnwidth}
	\includegraphics[width=\textwidth]{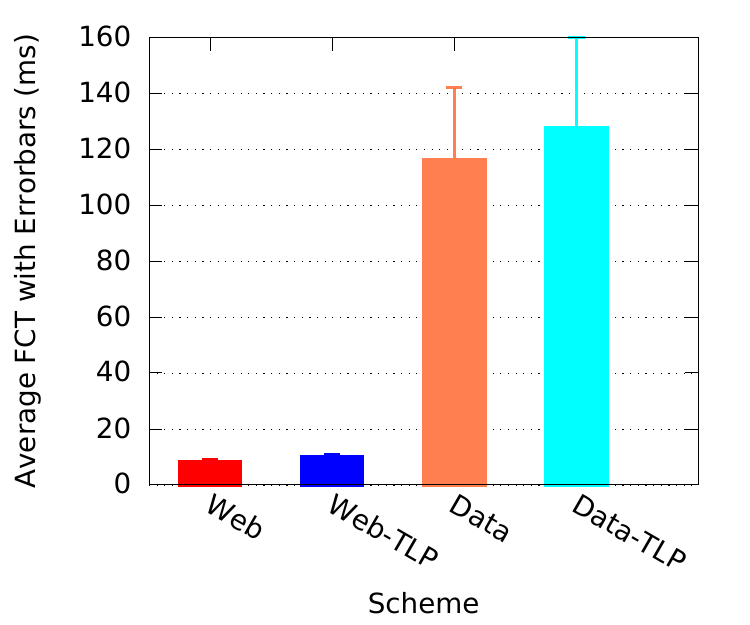}
	\caption{FCT of small flows}
	\label{fig:tlpnotlp}
	\end{subfigure}
	\caption{(a) shows the CDF of CWND at the time of the transmission of the lost packet. (b) TLP and NO-TLP FCT for Websearch and Datamining workloads}
	\label{fig:cwndandtlp}
\end{figure}

Analyzing these results, we can draw the following conclusions: Fig.~\ref{fig:analysis-frsize} suggests that FRR loss size is distributed over the range of packets with a positive skewness towards the first few fractions of the window (i.e., probability of losing more than 30\% of the window is insignificant). However, Fig.~\ref{fig:analysis-rtosize} shows that this is not the case for RTO, which seems well distributed with positive skewness towards the tail end of the window (i.e., the probability of losing more than 30\% of the window is significant). Also, we can see that there are only a few RTOs far away from the tail; these represent lost packets within the same congestion window. Fig.~\ref{fig:analysis-frpos} points out that losses at the tail of the window occur with higher frequency for RTO events, however in the case of FRR, the $Cwnd$ is relatively large enough for TCP to receive a sufficient number of duplicate Acks, which allows for Fast-Recovery. Similarly, Fig.~\ref{fig:analysis-rtopos} clearly shows a similar trend with higher frequency at the tail, however, in this case, $Cwnd$ is relatively small and hence, contains less in-flight packets to allow for FRR, and eventually, RTO recovery occurs. To elaborate, we see in Fig.~\ref{fig:cwndsize} that typically $Cwnd$ for the flows with segments that experience RTO is smaller than that of those that recover via FRR.

Finally, we also show the ineffectiveness of the TLP mechanism cited above \cite{TLP} in recovering tail losses in Fig.~\ref{fig:tlpnotlp}. The figure shows that TLP mechanism is not effective, and due to its additional overhead, it may even increase the FCT for small flows. 

In data-centers, the size of the pipeline is small: typically, with an RTT of 100us, a link of 1Gbps (respect. 10Gbps) can accommodate 8.3 packets (respect. 83 packets). In conjunction with shallow buffered switches, the nominal TCP fair share during TCP-incast barely exceeds one packet per-flow, and hence the occurrence of RTO is highly likely. This phenomenon highlights how TCP's performance can be degraded when operating in small windows regime in a small buffer with high-bandwidth low-delay environments like data-centers. The effect on the FCT is more severe for small, time-sensitive flows, that generally last only a few RTTs, but that are compelled to wait for 2 to 4 orders of magnitude extra time due to the $RTO_{min}$ rule.

\section{System Design and Implementation}
\label{sec:system}

T-RACKs design is based on the following observation: \emph{packet losses are inevitable in TCP.  So the key to reducing the long latency and jitter is not to try and avoid losses completely but instead try to avoid long waiting after the losses occur}. In achieving this, T-RACKS must also:
\begin{inparaenum}[(R1)]
\item improve the FCT of latency-sensitive applications by expediting the transmissions of small flows' packets.
\item be friendly to throughput-sensitive large flows (i.e., it must not sensibly degrade the throughput of large flows to satisfy the delay requirements of latency-sensitive flows).
\item be compatible with all existing TCP versions (i.e., it must not impose modifications inside the virtual machines, and if any are needed, they shall be in the hypervisors, which are fully under the control of the data center operator. It also must not require any extra special hardware).
\item Finally, the mechanism must be simple enough to be easily deployable in real data-centers.
\end{inparaenum}

In this perspective, T-RACKs actively infers packet losses by monitoring (in the hypervisor) per-flow TCP ACK numbers and proactively triggers the FRR mechanism of TCP to take action whenever and RTO is detected to be likely to take place. The goal is to help small TCP flows that would otherwise experience a timeout, recover fast via the FRR instead of waiting for TCP's $RTO_{min}$. The proposed mechanisms intervene when only the loss is almost certain, leading to a significant improvement of recovery times, and hence the FCT. T-RACKS design derives from the following arguments:
\begin{inparaenum}[\itshape i) \upshape]
\item all TCP versions adopt the FRR mechanism as a way to detect and recover from losses fast. So, if the FRR mechanism can be forced into action by the hypervisor, regardless of the nature of the loss, the resulting system would be transparent to the TCP protocol in the VM and would require no changes to TCP in the VM;
\item TCP relies on a small number of duplicate ACKs to activate FRR; however, in the majority of cases (especially for short-flows), there aren't enough packets in flight to trigger duplicate ACKs. To achieve this, we propose to use ``spoofed''  TCP ACK signaling from the hypervisor to the VM. In this perspective, the hypervisor maintains a per-flow timer $\beta = \alpha * RTT + rand(RTT)$ to wait for the ACKs before it triggers FRR with spoofed duplicate ACKs.
\end{inparaenum}

Note that our idea is similar in spirit to the so-called TCP SNOOP protocol~\cite{Balakrishnan1995}, which retransmits lost segments on behalf of the communicating end-points to filter out bit-errors in low-speed wireless networks. As such, TCP SNOOP also could be applied in data-centers. However, it is expensive to implement, as it requires buffering all sent segments at the lower layers (e.g., link-layer or hypervisor), which requires an ample buffer space in data-centers. T-RACKs, in contrast, triggers the retransmissions from the actual TCP protocol in the VM instead of buffering and retransmitting the packets itself. It requires no packet buffering at all; it only relies on memorizing a few state variables from the last segment, and the final ACK received of each flow.

\subsection{T-RACKs Algorithm}

\begin{algorithm}[!ht]%
\caption{T-RACKs Packet Processing}
\label{algo:rack1}
\small
\tcc{Initialization}
Create an in-memory flow cache pool\; \label{line:1}
Create flow table and reset flow information\;
Initialize and insert NetFilter hooks (for a NetFilter implementation)\;

\KwIn{$\alpha$\: \# of RTTs to wait before retransmitting ACKs}
\KwIn{$\gamma$\: a threshold in bytes to stop tracking a flow as small}
\KwIn{$\phi$\: the dupACK threshold used by TCP flows}
\KwIn{$t$: the current local time counted in jiffies}

Define $x$: the exponential backoff counter  \label{line:5}

\Fn{Outgoing Packet Event Handler (Packet P)}{
         f=Hash(P)\; \label{line:7}
				\If{SYN(P) or !f.active}
				{
					Reset Flow (f)\; \label{line:8}
					Extract TCP options (i.e, TStamp, SACK, etc)\;
					Update the flow information and set f.active\;\label{line:11}
				}
			  \If{DATA(P)}
				{
				      	Update flow info (i.e., last seq{\#}, etc)\;\label{line:13}
				      	f.active\_time = now()\;\label{line:14}
				}
}
\Fn{Incoming Packet Event Handler (Packet P)}{
				\tcc{For ACKs: extract and update flow information from incoming header}
				 f=Hash(P);\\ \label{line:17}
				  \lIf{f.long\_lived} 
					{
						 return \label{line:18}
					}
				\If{ACK\_bit\_set(P)} 
				{

					Extract required values (e.g., seq\#, ack{\#}, etc)\; \label{line:20}
					\eIf{New ACK}  
					{
						Update flow entry and state information (e.g., RTT)\; \label{line:22}
						Update last seen ACK number from receiver\; \label{line:23}
						Reset f.dupAck\_Nr = 0\; \label{line:24}
						Reset f.ACK\_time = now()\; \label{line:25}
						\lIf {f.lastAckNo $\geq \gamma$}
						{
							f.long\_lived = true \label{line:26}
						}
					}
					{
						\If{Duplicate ACK} 
						{
							f.dupAck\_Nr = f.dupAck\_Nr + 1\; \label{line:28}
						 	\tcc{Drop extra dup-ACKs}
							\lIf{f.resent $> 0$}
							{
							    Drop Dup ACK\label{line:31}
							}
						}
					}
					Update TCP headers (i.e., TStamps, SACK, etc)\;\label{line:32}
				}
}
\end{algorithm}
\normalsize

\begin{algorithm}[ht]%
\caption{T-RACKs Timeout Handler}
\label{algo:rack2}
\small
Create and  initialize a timer to trigger every 1 ms\;\label{line:2:1}
\Fn{Timer Expiry Event Handler}{
						\For{Flow (f) $\in$ FlowTable}
						{
								$\beta=\alpha * f.RTT + rand(f.RTT))$\; \label{line:2:4}
								\lIf{!f.active or f.long\_lived}
								{
									Continue \label{line:2:5}
								}
								
								T = MAX(f.ACK\_time, f.active\_time)\;\label{line:2:6}
								\If{now() - T $\geq \beta$}
								{
									Resend last ACK $\#(\phi - f.dupAck\_Nr)$ times\;\label{line:2:8}
									Set f.resent\_time = now()\;\label{line:2:9}
									Set x = 2\;\label{line:2:10}
									Continue\;\label{line:2:11}
								}
								\If{now()-f.resent\_time $\geq (\beta \ll x)$ }
								{
										resend ACK one more time\;\label{line:2:13}
										x = x + 1\;\label{line:2:14}
										Continue\;\label{line:2:15}
								}
								\If{(now()-f.ACK\_time) $\geq$ $RTO_{min}$}
								{
										stop T-RACKs recovery\;\label{line:2:17}
										soft reset flow (f) recovery state\;\label{line:2:18}
										Continue\;\label{line:2:19}
								}
								\lIf{(now()-f.active\_time)$\geq$1}
								{
								      deactivate\_flow(f) \label{line:2:20}
								}
						}
}
\end{algorithm}
\normalsize

The T-RACKs algorithm consists broadly of three major functions: the first two are in charge of maintaining per-flow state information on the server (hypervisor) on arrival and departure of packets, shown in Algorithm~\ref{algo:rack1} and the third is a timer event handler described in Algorithm~\ref{algo:rack2}. In the initialization in (lines $\ref{line:1}-\ref{line:5}$) of Algorithm~\ref{algo:rack1}, an in-memory flow cache pool is created to track new flow arrivals. This approach speeds up flow objects creation. A hash-based flow table is created and manipulated via the Read-Copy-Update (RCU) synchronization mechanism to efficiently identify flow entries. Other parameters and variables are set in this step, as well. Before each TCP segment departure, T-RACKs performs the following actions:
\begin{inparaenum}[\itshape i) \upshape]
\item in line $\ref{line:7}$, the packet is hashed using its 4-tuple (source and destination IP addresses and port numbers), and the corresponding flow is identified;
\item in lines $\ref{line:8}-\ref{line:11}$, if this is an SYN packet or the flow entry is inactive (i.e., a new flow), the flow entry is reset then TCP header info and options are extracted to activate a new flow record; and
\item in lines $\ref{line:13}-\ref{line:14}$, if this is a Data packet, then the $last_sent$ sequence number and time of the flow are updated.
\end{inparaenum}

Next, upon each TCP ACK arrival, the algorithm  performs the following actions:
\begin{inparaenum}[\itshape i) \upshape]
\item in lines $\ref{line:17}-\ref{line:18}$, the flow entry is identified using its 4-tuple; if the flow is large, we ignore it as it does not undergo recovery via T-RACKs. By doing so, the complexity of the scheme is reduced;
\item in lines $\ref{line:22}-\ref{line:26}$ if the ACK sequence number acknowledges a new packet arrival, the last seen ACK sequence number and time is updated. The dupACK counter is reset. If the accumulated flow size exceeds a threshold $\gamma$ it is marked as a large flow (to be able to stop tracking it);
\item in lines $\ref{line:28}-\ref{line:31}$, if ACK number acknowledges an old packet (i.e., if this is a duplicate ACK), then we drop dupACKs if the flow is in recovery mode, or otherwise increment the number of dupACKs seen so far;
\item in line $\ref{line:32}$, we update the TCP headers information of the ACK if necessary. We discuss this part in more detail later in sec~\ref{subsec:practical}.
\end{inparaenum}

Algorithm \ref{algo:rack2} handles the periodic global timer expiry events and performs the following actions for all \textbf{active} \textbf{non-large} flows in the table. In a typical implementation this timer lasts 1 ms and is processed regularly with the OS clock timer interrupt (i.e., does not require the special high-resolution timers):
\begin{inparaenum}[\itshape i) \upshape]
\item in lines $\ref{line:2:4}-\ref{line:2:11}$, if no new ACK acknowledging new data has arrived for $\beta$ seconds since the last new ACK arrival, the flow is deemed to be likely to experience a timeout in the future. T-RACKs enters into action, spoofs an ACK using the last successfully received ACK sequence number, and sends it out to the sending process or VM residing on the same end-host. An exponential backoff mechanism is activated to account for various dupACK thresholds set by the sender TCP or OS.
\item In lines $\ref{line:2:13}-\ref{line:2:15}$, if with timer $\beta$ backed off by the number of retransmissions $x$ of the spoofed ACK, the flow still did not receive a new ACK, another spoofed ACK is created and sent out to the corresponding sender. To ensure T-RACKS is not sending spurious spoofed dupAcks, the algorithm backs-off exponentially; i.e., after each transmission of a spoofed Ack,  timer $\beta$ is doubled.
\item In lines $\ref{line:2:17}-\ref{line:2:19}$, if the backoff time approaches the $RTO_{min}$ (i.e., 200ms), we stop triggering Fast-Retransmit (by resetting the soft state) and letting the sender's TCP RTO timer handle the recovery of this segment.
\item In line $\ref{line:2:20}$, if the inactivity period exceeds 1 sec, flow (f) entry is hard reset.
\end{inparaenum}

\subsection{T-RACKs System Implementation}
\begin{figure}[!t]
	\centering
		\includegraphics[scale=1.5, width=\columnwidth]{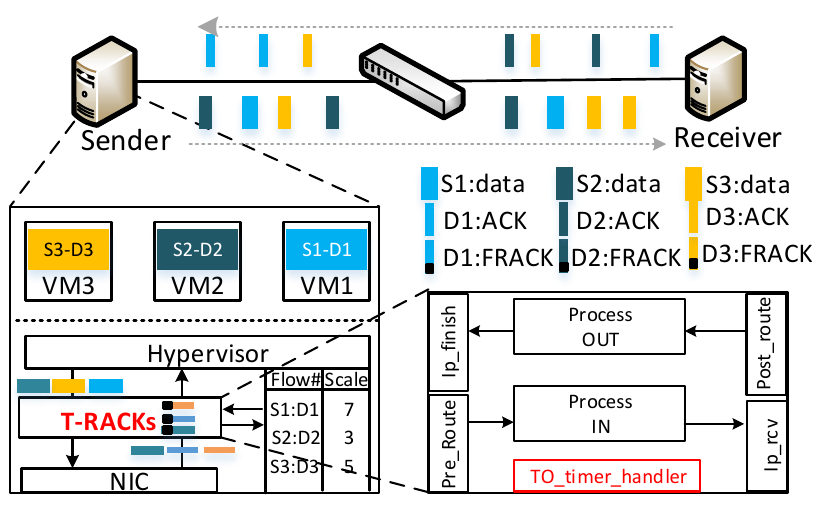}
	\caption{T-RACKs System: It consists of an end-host module that track TCP flows incoming ACKs and generates FAKE ACKs whenever a flow timesout}
	\label{fig:system}
\end{figure}

Algorithm~\ref{algo:rack1} relies on TCP header information of ACK packets to maintain per-flow TCP state information. In this paper, we only consider a lightweight end-host (hypervisor) shim-layer implementation to achieve this\footnote{We note that, in higher speed networks, T-RACKs could equally be implemented as a networking function on smartNICs}. This approach is perfectly feasible even for production data-centers, because the number of flows in a server in production data-centers has been reported to be small in general not exceeding 30-40 \cite{Alizadeh2010}. In addition, the number of flows tracked by T-RACKs is further reduced on average by only tracking small-sized flows, abandoning large flows whenever they grow to reach a certain size threshold. The deployment of T-RACKs in data-centers involves hashing the flows into a hash-based flow-table using the 4-tuples (i.e, SIP, DIP,  Sport and Dport) whenever SYNs packets arrive or a flow sends data after a long silence period. For instance, referring to Figure~\ref{fig:system}, when VM1 on the sender end-host established a connection with its peer (or destination) VM on the receiving end-host, a new flow entry  ($S2:D2$) is created in the table as shown in Fig.\ref{fig:system}. Also, not shown in the Algorithm code, flow entries are cleared from the table whenever a connection is closed (following the TCP connection tear down FIN/FIN-ACK) or after a pre-set inactivity time threshold is exceeded. The flow table could track many relevant TCP-related per-flow state information, however, for T-RACKs to perform the fast recovery function, it needs to track a minimal set of TCP state variables (including the highest ACK sequence number seen so far and the arrival time of the most recent ACK).

The T-RACKs system uses a flow table to store and update TCP flow information, including the last ACK number, the last sent sequence number, the corresponding times, the RTT for the flow measured using the TCP timestamp option, as well as the optional TCP Sack information\footnote{Note that if TCP Sack is active, TCP's response to duplicate ACKs is different from the standard behavior, therefore we need to take this into account to elicit a proper reaction.} for each outgoing TCP flow. T-RACKs intercepts the incoming ACKs and outgoing Data to update the current state of each tracked small flow. When packets are dropped by the network and the receiver receives enough out of sequence DATA to generate sufficient dupACKs (real ones), the loss is recovered via FRR from the VM without the intervention of T-RACKs. However, when the receiver fails to receive enough DATA to generate enough real dupACKs to trigger FRR, then T-RACKs intervenes after a timer (1ms) by sending spoofed dupACKs (or RACKs for retransmitted-ACKs) to the sender. Typically, the sender would receive enough dupACKs and RACKs to trigger FRR to retransmit the lost segment within a reasonable time, long before the TCP RTO is reached.

\subsection{Practical Aspects of T-RACKs System}
\label{subsec:practical}
\begin{figure}[!t]
	\centering
		\includegraphics{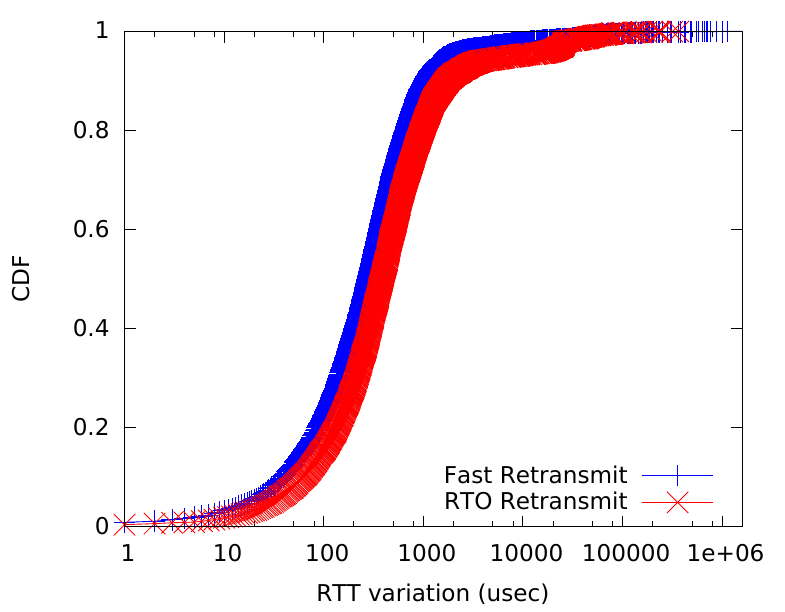}
	\caption{RTT variation between transmission time and time of retransmission(i.e $\Delta RTT=RTT_{rt} -RTT_t$)}
	\label{fig:RTTvar}
\end{figure}

\textbf{T-RACKs System:} is built upon a light-weight module at the hypervisor layer tracking a limited per-flow state. In the simplest case, it tracks TCP's identification 4-tuple, per-flow last ACK number, and the timestamp of the last non-dupACK packet. The system in spirit is similar to recent works in \cite{Ratcliff2016, He2016} that aim to enable virtualized congestion control in the hypervisor without cooperation from the tenant VM, however these approaches require fully-fledged TCP state information tracking and implementing full TCP finite-state machines in the hypervisor including packet queuing. On the other hand, T-RACKs tries to minimize the overhead by tracking the minimal amount of necessary information and implementing only a subset of the retransmission mechanism while letting the VM do the actual work of transmission and queuing.

\textbf{Complexity:} T-RACKs Complexity resides in its interception of ACKs to update the last seen ACK information. Since, it does not perform any computation on the ACK packets,  it does not add much to the load on the hosting server nor to the latency\footnote{ACKs is updated in some instances (e.g., to insert fake SACK block to signify a small gap in the SACKed numbers. Otherwise RACK packets would be ignored by TCP).}. This claim is supported by our observation in our experiments on our data-center. A hash-based table is used to track flow entries of \textbf{active small} flows. In the worst case, when hashes collide, a linear search is necessary within the linked-list. However, this worst case is rare due to the small number of flows originating from a given end-host. Typically, end-host CPUs can sustain rates of 60 Gbps of packet processing. Hence, the little processing required by the insertion of T-RACK state would not affect the TCP throughput. 

\textbf{Spurious retransmissions:} T-RACKs may possibly introduce spurious retransmissions making in-network congestion worse. However, this boils down to answering similar question when choosing the correct RTO value in TCP. For this purpose, we refer to a previous study \cite{Allman1999}, that mostly showed that even when a relatively bad RTT estimator is used, setting a relatively high minimum RTO, it can help avoid many of the spurious retransmissions in WAN transfers. This fact is supported by a subsequent study \cite{Zhang2001} that shows significant changes (or variance) in Internet delays. Recent works \cite{Guo2015, Mittal2015} show similar behavior within current data-centers. In our testbed, we observed noticeable variation in the measured RTT. To quantify this, we measured the difference in RTT values collected at the time of the first transmission of the packet and then at the time of fast retransmission or RTO retransmission. From the collected data, a considerably large variation, ranging from a few hundred microseconds at the $\approx$20th percentile to a thousand microseconds at the $\approx$80th percentile, is observed in the smoothed RTT at the TCP sockets, as shown in Fig.~\ref{fig:RTTvar}. These variations can be mainly attributed to the beginning of some heavy background traffic, imbalance introduced by load balancing, or VM migrations, and so on. We note and agree with the aforementioned works that observed packet delays may not be mathematically nor stochastically steady. Hence T-RACKs ACK RTO ($\beta$) calculation shown in Algorithm~\ref{algo:rack1} strikes a balance between rapid retransmission and the risk of causing spurious retransmission.

\textbf{T-RACKs RTO $\beta$: } in most of our experiments and simulations, we choose a value for ACK RTO ($\beta$) to be ($\geq$ 10) times the dominant measured RTT in the data-center. We believe, and the results show that this value achieves a good tradeoff between not having many of the spurious retransmissions and, at the same time not being too late in recovering from losses. We further adopt the well-know exponential back-off mechanism for subsequent ACK RTO ($\beta$) calculations until either the loss is recovered or TCP's default RTO (i.e., $RTO_{min}$) is close enough to be reached.

\textbf{Synchronization of retransmissions: }
Since T-RACKs relies on a timer for ACK recovery, such timer may result in the synchronization of retransmissions from different VMs on different hosts leading into incast-like congestion. We studied this behavior in a simulation, and the results show repeated losses due to possible synchronized retransmissions. A viable solution to de-synchronize such flows is to introduce some randomness in the ACK RTO, ultimately resulting in fewer flows experiencing repeated timeouts. We adopted this approach and added a random delay in the calculation of the RTO $\beta$, as shown above in the algorithms.

\textbf{TCP Header manipulation:} TCP does not accept any packet with an inconsistent timestamp, hence the timestamps are updated per ACK arrival with the local $jiffies$ variable to keep the consistency of timestamps whenever RACKs are sent. For SACK-enabled TCP, fake SACK block information needs to be inserted for incoming ACKs (with no SACK blocks in TCP header) to indicate a small gap equal to the minimum segment size (i.e., 40 Bytes) after the last successfully acknowledged data.

\textbf{Security Concerns: } during FRR, to be able to maintain its flight size and avoid timeout, TCP inflates the window artificially by 1MSS for each received dupAck. This can be exploited to launch ACK spoofing attack \cite{Savage1999} on the senders. RFC 5681 released in 2009 addressed this particular attack and proposed implementing Nonce and Nonce-Reply as a way of verifying the source of dupACKs. However, such a solution would require the introduction of extra TCP headers prohibiting its deployment in real TCP implementations. In T-RACKs, we address such attack by dropping dupACKS whenever the ACK timer expires when entering a recovery state. This approach is adopted to disable $Cwnd$ artificial inflation during recovery and, at the same time, prevents external ACK spoofing (other than RACKs). Worth mentioning also is that RACKs are generated from the hypervisor layer, which is under the control of the trusted data-center operator.

\textbf{TCP semantics: } is conceptually violated since dupACKs should reflect packets following the lost one being received successfully. However, according to RFC 5681, the network could replicate packets, and hence the RACKs could be treated as replicated packets from within the network.

\begin{figure}[t]
	\centering
		\includegraphics[scale=1]{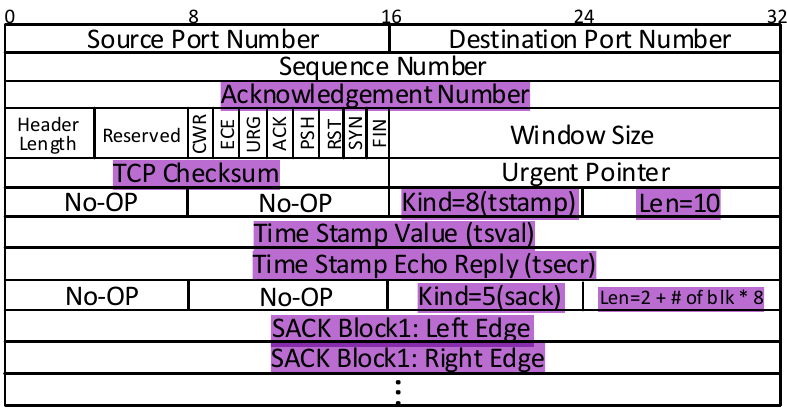}
	\caption{TCP headers manipulated by T-RACKs system}
	\label{fig:header}
\end{figure}

\textbf{TCP Header manipulation:} Fig.~\ref{fig:header} shows TCP headers manipulated by T-RACKs module, the module tracks and uses the highlighted TCP headers for all incoming ACK packets using RCU flow entries.

\section{Simulation Analysis}
\label{sec:sim}
In this section, we study the performance of T-RACKs to verify if it can achieve its goals in small-scale and large-scale simulation scenarios. We first conduct a number of packet-level simulations using ns2 and compared T-RACKs performance to the state-of-the-art schemes. (For brevity, we refer to T-RACKs as RACK in the figures.)

\subsection{Simulations in a Dumbell Topology}
To study how TCP behaves in response to packet losses and how likely it may recover quickly with the help of T-RACKs, we conducted several packet-level simulation experiments to covers a large variety of TCP and AQM combinations. We also conducted simulation experiments using the congestion control mechanisms code imported from the Linux kernel (i.e., New Reno and Cubic). We use ns2 version 2.35 \cite{NS2}, which we have extended with T-RACKs mechanism inserted as a connector between nodes and their link in the topology setup. Besides, we patched ns2 using the publicly available DCTCP patch.  We use in our simulation experiments link speeds of 1 Gbps for sending stations, a bottleneck link of 1 Gb/s, low RTT of 100 $\mu$s, the default TCP $RTO_{min}$ of 200 ms and TCP initial window of 10 MSS. We use a rooted tree topology with a single bottleneck at the destination and run the experiments for 15 sec. The buffer size of the bottleneck link is set to 100 packets, which is more than the bandwidth-delay product in all cases. 
We first designed two simulation scenarios: 
\begin{enumerate}
\item CASE 1: Small flows.
\item CASE 2: Large flows coexisting with small flows.
\end{enumerate} 

In CASE 2, the ratio of small flows to large flows is set to 3, and large flows send data for the whole duration of the experiment. In both cases, each small flow sends a 14.6KB file (i.e., 10 MSS) until it completes its transmission. Small flows start with a random inter-arrival time that is drawn from an exponential distribution with a mean equal to the transmission time of one packet; this allows us to create clusters of small flows that start almost simultaneously, to emulate incast traffic. This process is repeated once every 3 sec, which gives five rounds of incast traffic,  during the simulation. In each round, we draw the order of the servers generating the flows according to a uniform distribution. We study packet losses, the likelihood of fast recovery, and recovery time. We first study TCP newReno with Droptail,  RED-ECN, and Random Drop AQMs, and DCTCP covering the most common TCP and AQM settings.

\begin{figure}[t]
\captionsetup[subfigure]{justification=centering}
\centering
  \begin{subfigure}[ht]{0.48\columnwidth}
           \includegraphics[width=\columnwidth]{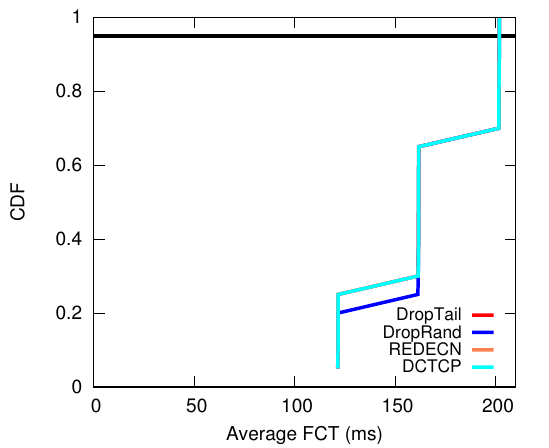}
	\caption{Without T-RACKs}
	\label{fig:20flows1-norack}
       \end{subfigure}
    \begin{subfigure}[ht]{0.48\columnwidth}
           \includegraphics[width=\columnwidth]{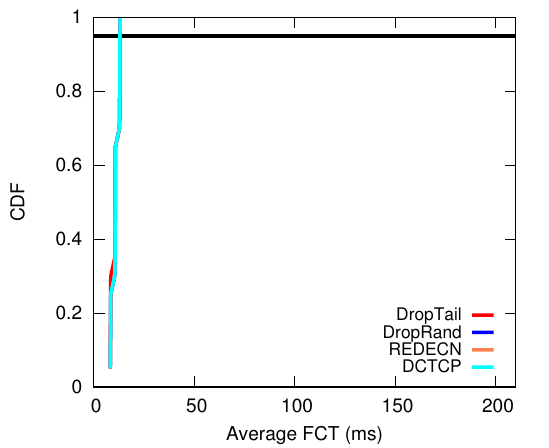}
	\caption{With T-RACKs}
	\label{fig:20flows1-rack}
       \end{subfigure}
\caption{CDF of average FCT for CASE 1 in 20 flows scenario}
\label{fig:simavgfct1}
\end{figure}

\begin{figure}[t]
\captionsetup[subfigure]{justification=centering}
\centering
\begin{subfigure}[ht]{0.48\columnwidth}
       \includegraphics[width=\columnwidth]{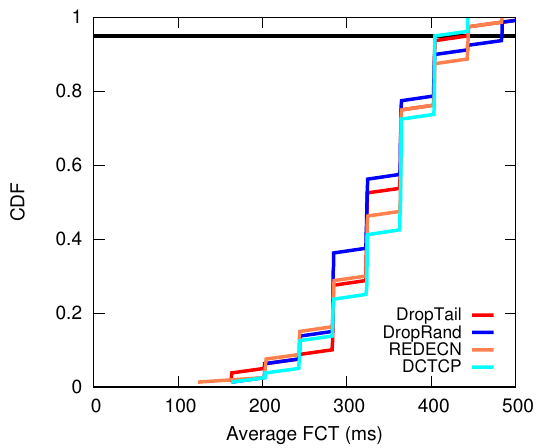}
          \caption{Without T-RACKs}
        \label{fig:80flows1-norack}
      \end{subfigure}				
    \begin{subfigure}[ht]{0.48\columnwidth}
       \includegraphics[width=\columnwidth]{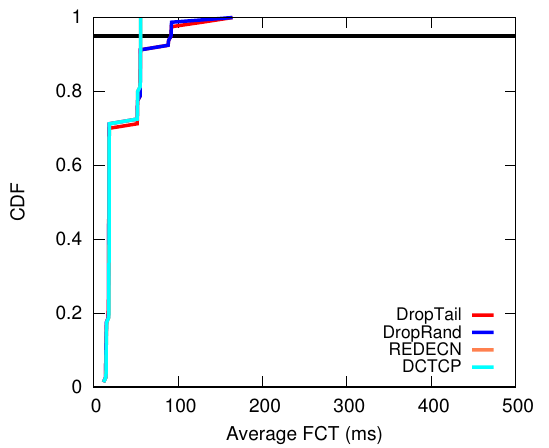}
          \caption{With T-RACKs}
        \label{fig:80flows1-rack}
      \end{subfigure}				
\caption{CDF of average FCT for CASE 1 in 80 flows scenario}
\label{fig:simavgfct2}
\end{figure}

First, we run the experiments without T-RACKs for 20 and 80 flow traffic scenarios generated according to CASE 1 (i.e., containing only small flows). Figures~\ref{fig:20flows1-norack} ~and~\ref{fig:80flows1-norack} show the average FCT for different schemes. It appears that the FCT is significantly high for all-schemes exceeding 200ms for most flows (more than $\approx$35$\%$ and $\approx$95$\%$ for 20 and 80 flows, respectively). This result indicates that most flows are experiencing RTOs. We repeat the same simulation, enabling T-RACKs on the end-hosts, to mitigate the RTOs. Figures~\ref{fig:20flows1-rack} ~and~\ref{fig:80flows1-rack} show the FCT for different schemes with T-RACKs for 20 and 80 flows scenario in CASE 1. The results show significant improvement in the average FCT for the two scenarios for all schemes. Specifically, it could reduce the average FCT range significantly for both 20 and 80 flows scenarios. In the 20 flows scenario, at $95^{th}\%$ (as highlighted by the horizontal black line), the reduction in average FCT is up to $\approx$15 times (i.e., from 200ms down to 13ms). In the 80 flows scenario, it reduces the FCT  at $95^{th}\%$ by $\approx$3 times for DropTail and DropRand,  by $\approx$8 times for RED-ECN and DCTCP, respectively.  %

\begin{figure}[t]
\captionsetup[subfigure]{justification=centering}
\centering
  \begin{subfigure}[ht]{0.48\columnwidth}
           \includegraphics[width=\columnwidth]{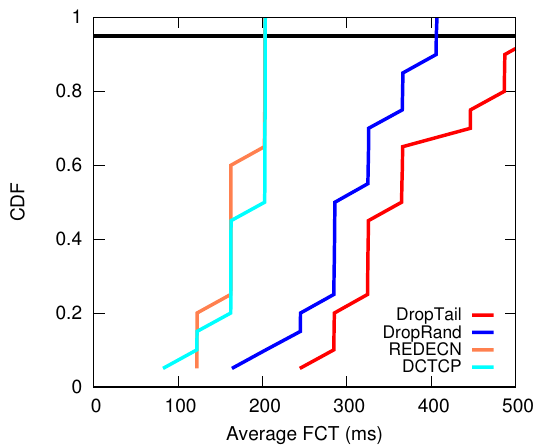}
	\caption{Without T-RACKs}
	\label{fig:20flows2-norack}
       \end{subfigure}
    \begin{subfigure}[ht]{0.48\columnwidth}
       \includegraphics[width=\columnwidth]{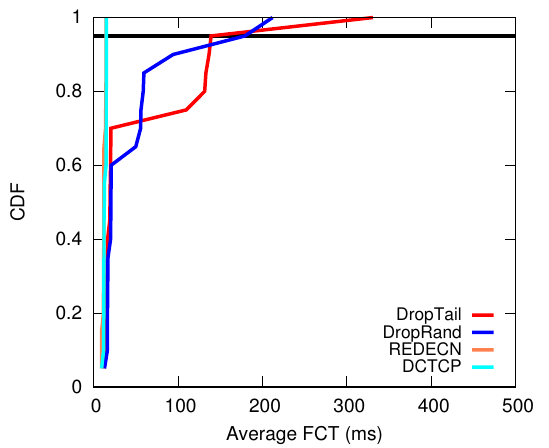}
          \caption{With T-RACKs}
        \label{fig:20flows2-rack}
      \end{subfigure}					
\caption{CDF of average FCT for CASE 2 in 20 flows scenario}
\label{fig:simavgfct3}
\end{figure}

\begin{figure}[t]
\captionsetup[subfigure]{justification=centering}
\centering
       \begin{subfigure}[ht]{0.48\columnwidth}
           \includegraphics[width=\columnwidth]{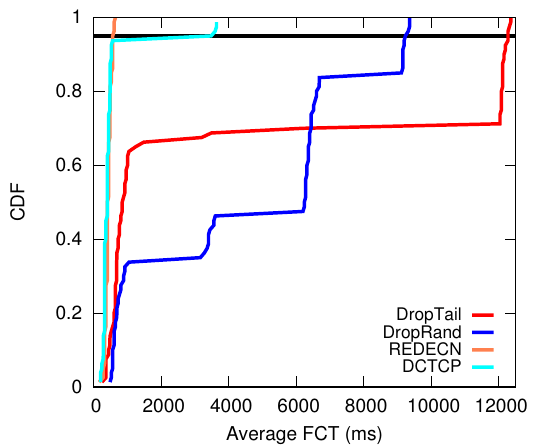}
	\caption{Without T-RACKs}
	\label{fig:80flows2-norack}
       \end{subfigure}
    \begin{subfigure}[ht]{0.48\columnwidth}
       \includegraphics[width=\columnwidth]{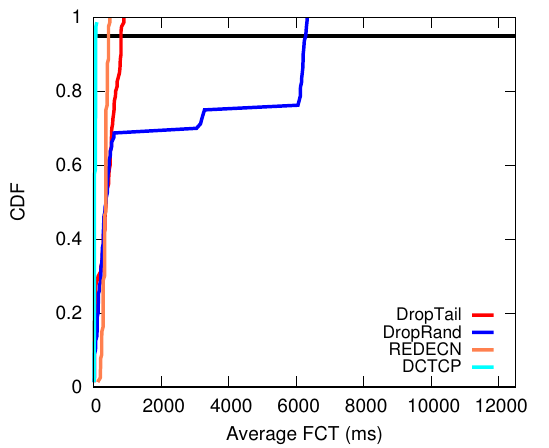}
          \caption{With T-RACKs}
        \label{fig:80flows2-rack}
      \end{subfigure}							
\caption{CDF of average FCT for CASE 2 in 80 flows scenario}
\label{fig:simavgfct3bis}
\end{figure}

Now we perform simulations for CASE 2, which introduces large flows (i.e., background traffic) to CASE 1. This case is use to study T-RACKs under stress. Figure~\ref{fig:20flows2-norack} and Figure~\ref{fig:20flows2-rack} show the results without and with T-RACKs for 20 flows scenario. The results show a significant reduction in the average FCT of small flows. In the 20 flows scenario, T-RACKs reduces the average FCT on $95^{th}$\% by $\approx$1.6X, $\approx$1.9X, $\approx$14X and $\approx$14X for DropTail, DropRand, RED-ECN, and DCTCP, respectively. Figure~\ref{fig:80flows2-norack} and Figure~\ref{fig:80flows2-rack} show the results without and with T-RACKs for the 80 flows scenario. In the 80 flows scenario, T-RACKs can further improve the performance, and it manages to reduce the average FCT at the $95^{th}$\% by $\approx$14X,  $\approx$1.5X, $\approx$1.4X and $\approx$37X for DropTail, DropRand, RED-ECN, and DCTCP, respectively. These results show that T-RACKs improves the performance more in heavy load scenarios with background traffic.

\begin{figure}[t]
\captionsetup[subfigure]{justification=centering}
\centering
  \begin{subfigure}[ht]{0.48\columnwidth}
           \includegraphics[width=\columnwidth]{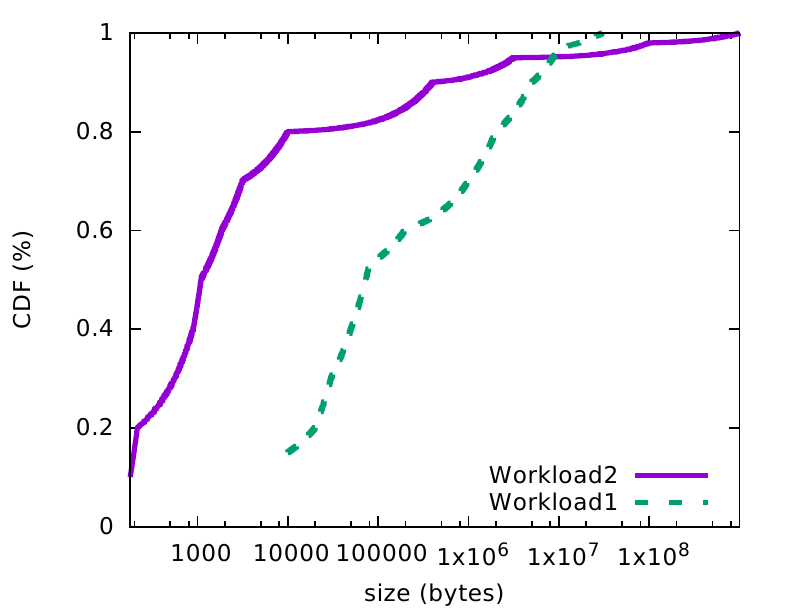}
	\caption{CDF of flow size}
	\label{fig:simflowsize}
       \end{subfigure}
    \begin{subfigure}[ht]{0.48\columnwidth}
       \includegraphics[width=\columnwidth]{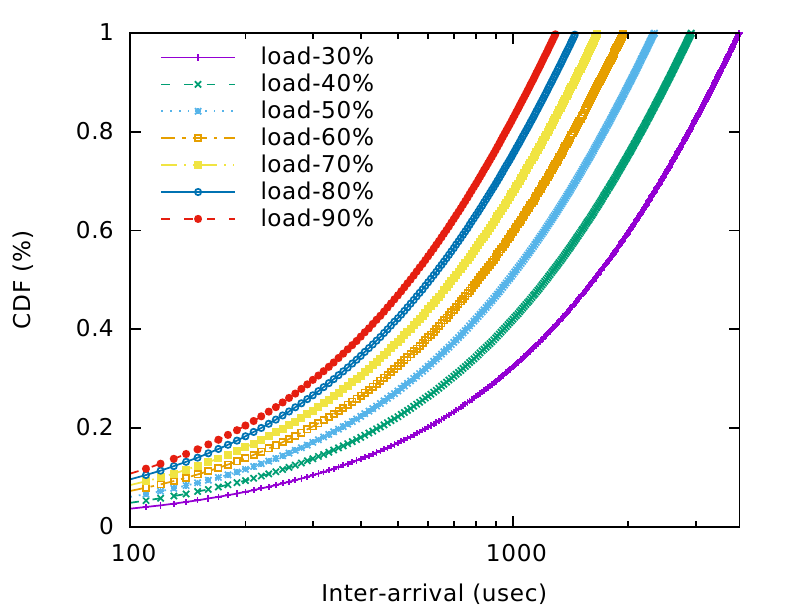}
          \caption{CDF of inter-arrival time}
        \label{fig:siminterarrival}
      \end{subfigure}
\caption{Flow characteristics: (a) Actual Flow size distribution (b) Inter-arrival times for various network load}
\label{fig:sizeinter}
\end{figure}

\subsection{Simulations in Data-center Topology}

End-host based schemes are assumed by default to be scalable and to verify this, we experiment with T-RACKs on a larger scale with varying workloads and flow size distributions. 

For this purpose, we conduct another packet-level simulation using a spine-leaf topology with nine leaf switches and four spine switches using links of 10 Gbps for end-hosts and an over-subscription ratio of 5\footnote{the typical over-subscription ratio in current production data-centers is in range of 3 to over 20}. We again examine scenarios with TCP-NewReno, TCP-ECN, and DCTCP operating along with Droptail, RED, and DCTCP AQMs, respectively.  We use a per-hop link delay of 50 $\mu$s, TCP is configured with the default TCP $RTO_{min}$ of 200 ms and an initial window of 10 MSS. Persistent connections are used for successive requests. Finally, buffer sizes on all the links are set to be equal to the bandwidth-delay product between end-points within one physical rack. 

The flow size distribution for workload 1 (which represents Websearch flow sizes distribution \cite{Alizadeh2010}) and workload 2 (which represents datamining flow sizes distribution~\cite{Greenberg2009}) are shown in Fig~\ref{fig:simflowsize} which capture a wide range of flow sizes. The flows are generated randomly from any source host to any other destination host with the arrivals following a Poisson process with various average flow arrival rates to simulate different network loads. Fig~\ref{fig:siminterarrival} shows the inter-arrival times distribution for various traffic loads ranging from $30\%$ to $90\%$. 

We report the average FCT for small flows and for all flows\footnote{Flow sizes in the range [0-100KB] are considered to be small, flow sizes in the range [100KB - 10MB] are considered to be medium and flows sizes from 10MB and upwards are considered to be large flows.} as well as the total number of timeout events in each case. In the simulation, the T-RACKs threshold $\gamma$ is set to infinity (i.e., all flows are tracked including large flows). The T-RACKs RTO, the timeout to trigger spoofed dupACKs, is set to 10 times the measured RTT in this experiment. That means if an ACK with the expected sequence number is not delivered within 10 RTTs then RACK packets are spoofed to trigger the FRR.

\begin{figure*}[ht]
\captionsetup[subfigure]{justification=centering}
\centering
       \begin{subfigure}[ht]{0.32\textwidth}
             \includegraphics[width=\textwidth]{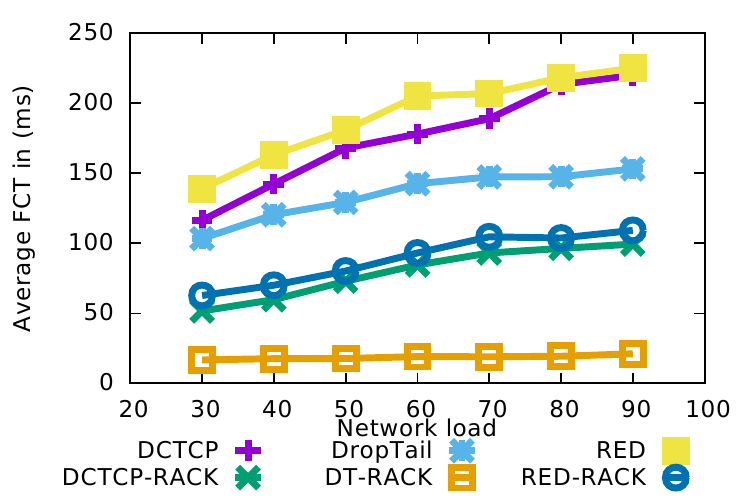}
      \caption{Small Flows: Average FCT}
              \label{fig:smallavgfctdata}
       \end{subfigure}
       \hfill
       \begin{subfigure}[ht]{0.32\textwidth}
             \includegraphics[width=\textwidth]{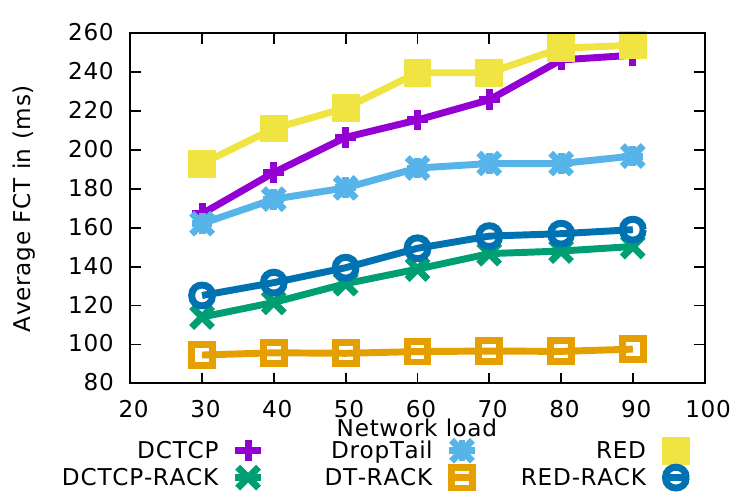}
             \caption{All Flows: Average FCT}
              \label{fig:allavgfctdata}
       \end{subfigure}
	\hfill
	\begin{subfigure}[ht]{0.32\textwidth}
           \includegraphics[width=\textwidth]{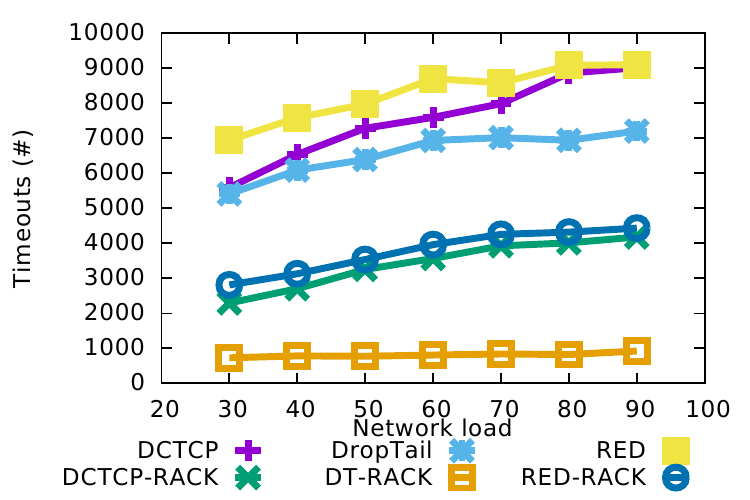}
					\caption{All Flows: Number of RTOs}
     \label{fig:alltodata}
       \end{subfigure}
    \\
    \begin{subfigure}[ht]{0.32\textwidth}
             \includegraphics[width=\textwidth]{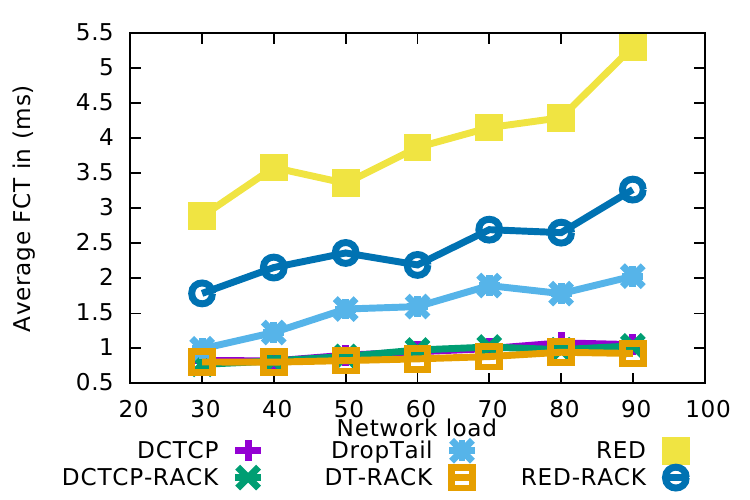}
      \caption{Small Flows: Average FCT}
              \label{fig:smallavgfctdata}
       \end{subfigure}
       \hfill
       \begin{subfigure}[ht]{0.32\textwidth}
             \includegraphics[width=\textwidth]{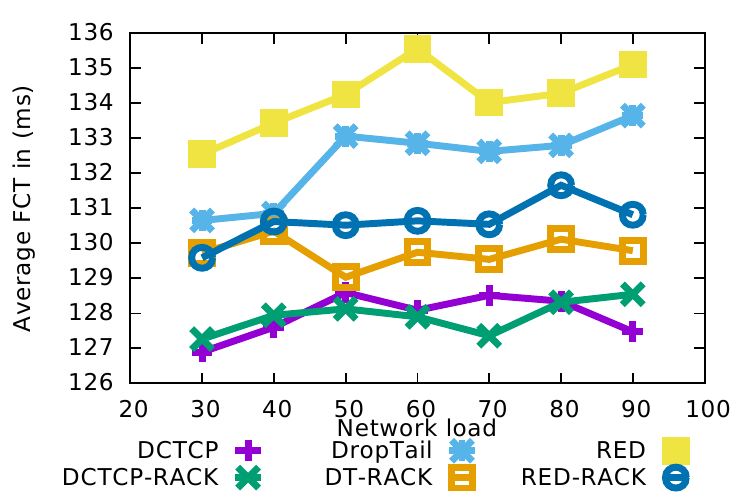}
             \caption{All Flows: Average FCT}
              \label{fig:allavgfctdata}
       \end{subfigure}
	\hfill
	\begin{subfigure}[ht]{0.32\textwidth}
           \includegraphics[width=\textwidth]{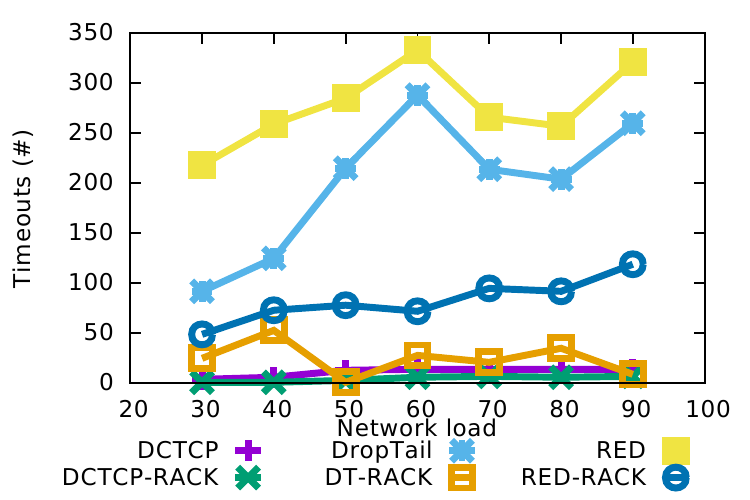}
					\caption{All Flows: Number of RTOs}
     \label{fig:alltodata}
       \end{subfigure}
      \caption{Performance with network load in (30$\%$, 90$\%$)  for workload 1  (top) and workload 2 (bottom)}
	\label{fig:simexp1}
  \vspace{-1.5em}
\end{figure*}

The average FCT for small and all flows, as well as the total number of timeouts experienced by all flows for workloads 1 and workload 2 are shown in Fig.~\ref{fig:simexp1}. Note that for both workloads, the FCT of small flows is dramatically degraded when they experience a timeout regardless of the TCP congestion control version or AQM mechanism in operation. In contrast, when T-RACKs is activated, it helps small flows the most by improving their FCT as a by-product of reducing the number of timeouts they experience. We also note that the overall FCT decreases for all flows for two reasons: 1) the threshold $\gamma$ enables all flows to benefit from T-RACK and 2) small flows finish quicker leaving network resources for larger ones. We notice that for workload 2, with almost 80$\%$ of the flows being less than 10KB and hence experiencing lesser timeout events overall, DCTCP can improve the FCT. This improvement is because DCTCP's ability to regulate the persistent queue length (i.e., there are few large flows to fill the buffer, unlike workload 1).

\subsection{Sensitivity to Choice of T-RACKs RTO}

\begin{figure*}[t]
\captionsetup[subfigure]{justification=centering}
\centering

  \begin{subfigure}[ht]{0.32\textwidth}
    \includegraphics[width=\textwidth]{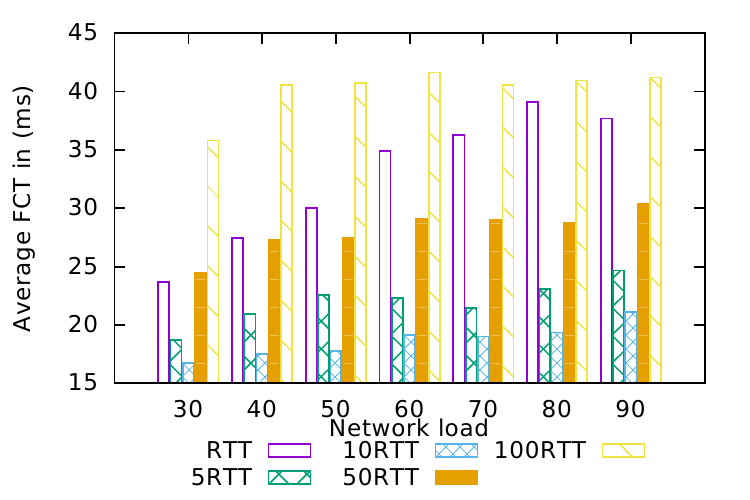}
	\caption{Small Flows: DropTail}
	\label{fig:droptailsmallrto}
       \end{subfigure}     
      \hfill
     \begin{subfigure}[ht]{0.32\textwidth}
     \includegraphics[width=\textwidth]{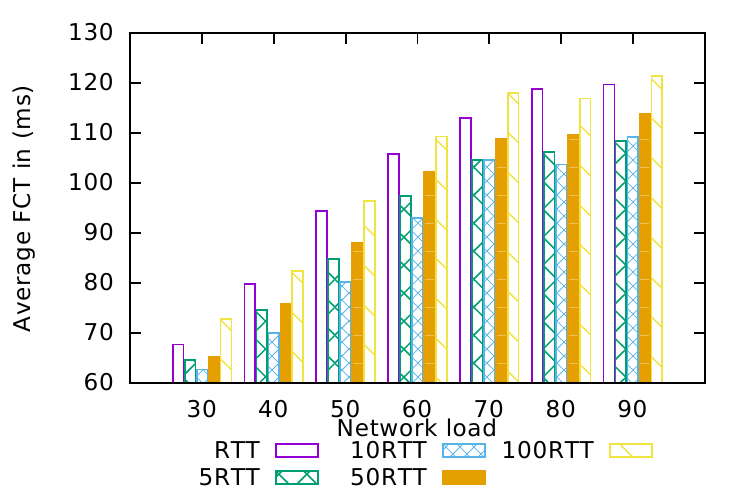}
	\caption{Small Flows: RED-ECN}
	\label{fig:redsmallrto}
       \end{subfigure}  
     \hfill
    \begin{subfigure}[ht]{0.32\textwidth}
    \includegraphics[width=\textwidth]{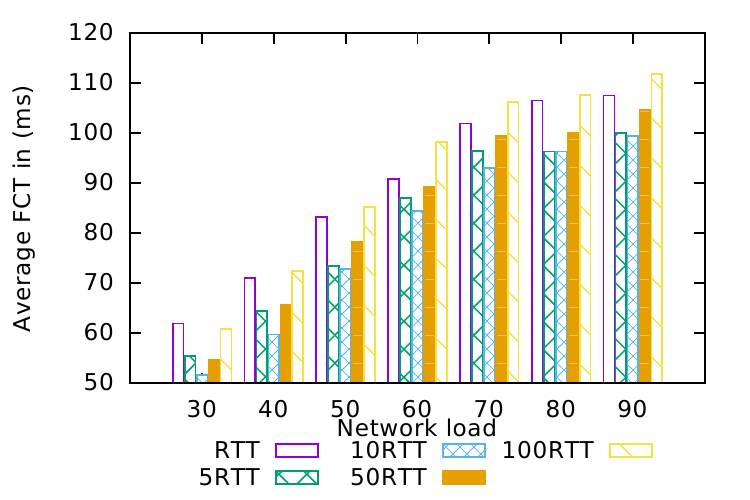}
      \caption{Small Flows: DCTCP}
        \label{fig:dctcpsmallrto}
      \end{subfigure}
      
\caption{The same Websearch scenario as above but using DropTail AQM and $\alpha$ is varied from 1 RTT to 100 RTTs.}
\label{fig:varyrto}
  \vspace{-1.5em}

\end{figure*}

In this experiment, we study the sensitivity of T-RACKs to the preset RACK RTO value. For this purpose, we repeat the last simulation by varying the value of the RTT multiplicative factor $\alpha$ in the set [1, 5, 10, 50, 100]. We report the average FCT of small flows and all flows in each case for DropTail, RED, and DCTCP in Figure~\ref{fig:varyrto} for various loads. From the figure, we can see that the FCT is greatly affected by the choice of parameter $\alpha$. Small values of $\alpha$ (i.e., 1 and 5) show a relatively large FCT compared to the RTT, which indicates that they tend to cause too many spurious retransmissions that exacerbate congestion in the network. On the other hand, excessively large values for $\alpha$ (i.e., 50 and 100) tend to be too conservative and result in TCP flows recovering later than they could. We can see in the three figures for all loads a minimum FCT is achieved at or near a RACK RTO of 10 RTTs.

\section{Linux Kernel Implementation}
\label{sec:imp}

\begin{figure*}[ht]
\captionsetup[subfigure]{justification=centering}
\centering
	\begin{subfigure}[ht]{0.32\textwidth}
             \includegraphics[width=\textwidth]{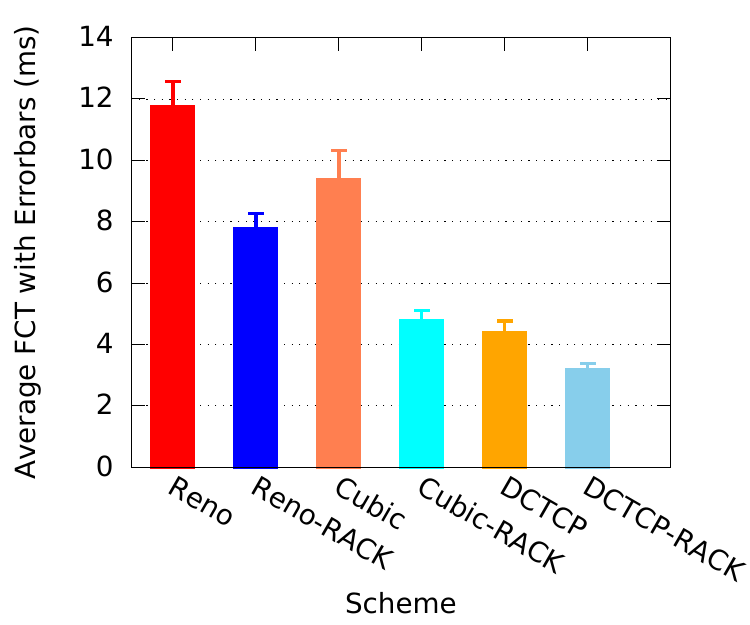}
      \caption{Small Flows: Average with Errorbar}
              \label{fig:smallavgfct1}
       \end{subfigure}
       \hfill
       \begin{subfigure}[ht]{0.32\textwidth}
             \includegraphics[width=\textwidth]{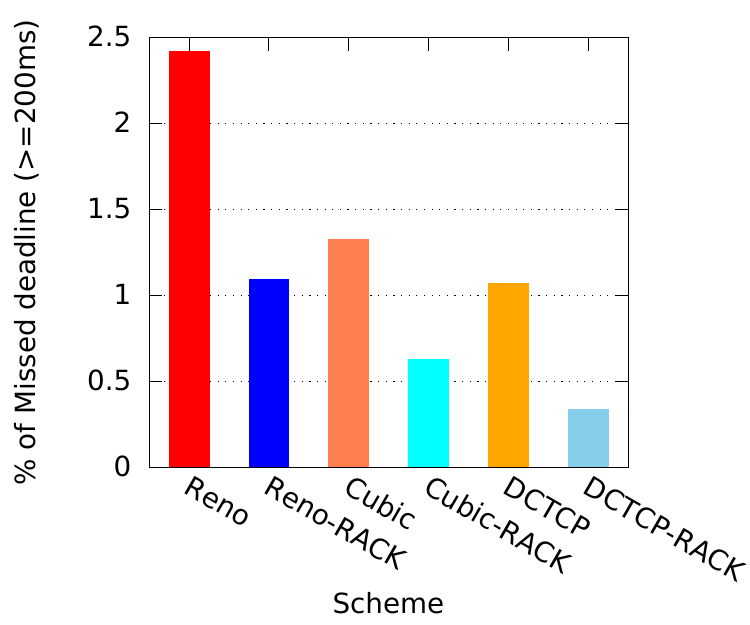}
      \caption{Small Flows: Missed Deadlines}
              \label{fig:smallmiss1}
       \end{subfigure}
	\hfill
	\begin{subfigure}[ht]{0.32\textwidth}
           \includegraphics[width=\textwidth]{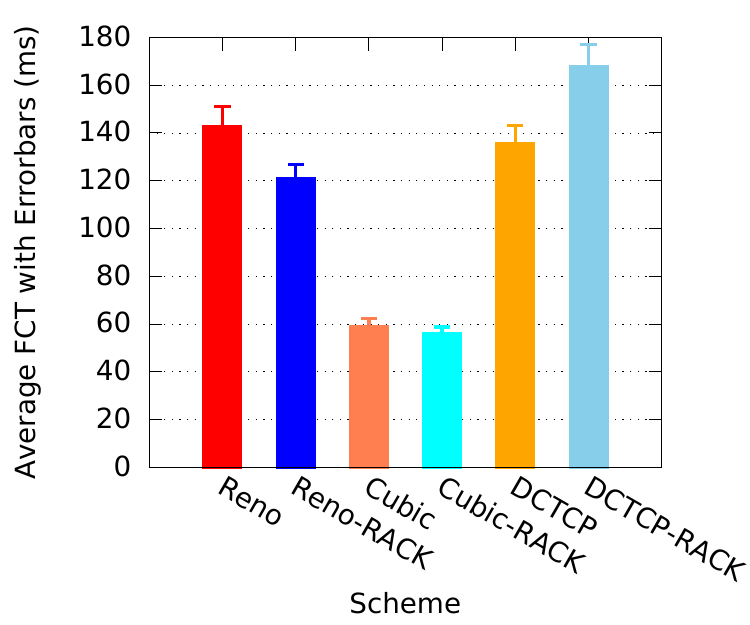}
	\caption{All Flows: Average with Errorbar}
     \label{fig:allavgfct1}
       \end{subfigure}
       \\
       \begin{subfigure}[ht]{0.32\textwidth}
             \includegraphics[width=\textwidth]{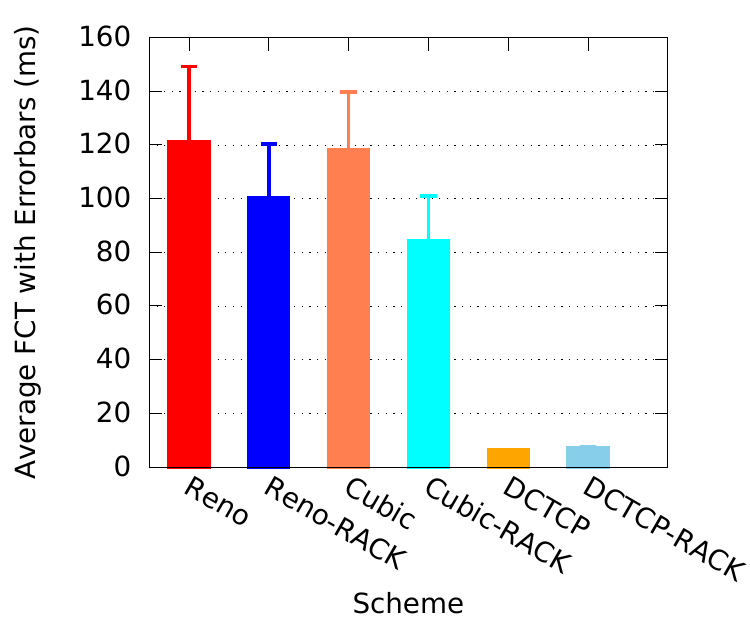}
      \caption{Small Flows: Average (Datamining)}
              \label{fig:smallavgfctdatmin1}
       \end{subfigure}
       \hfill
       \begin{subfigure}[ht]{0.32\textwidth}
             \includegraphics[width=\textwidth]{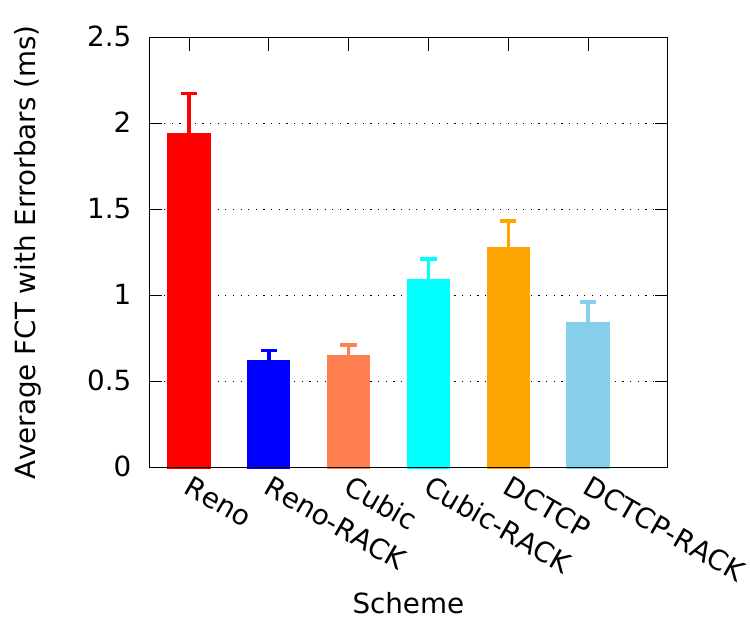}
             \caption{Small Flows: Average (Educational)}
              \label{fig:smallavgfctedu1}
       \end{subfigure}
	\hfill
	\begin{subfigure}[ht]{0.32\textwidth}
           \includegraphics[width=\textwidth]{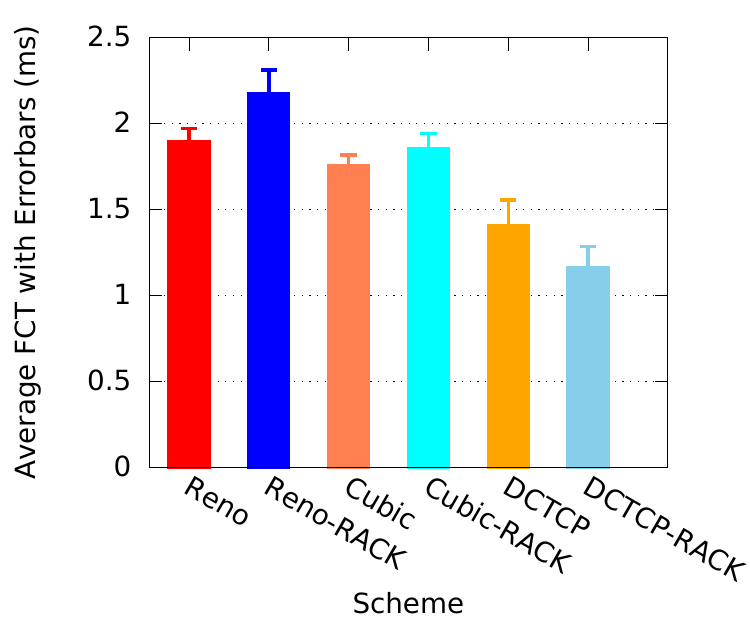}
	    \caption{Small Flows: Average (Private DC)}
     \label{fig:smallavgfctprv1}
       \end{subfigure}

\caption{Performance metrics of one-to-all scenario without any background traffic for (a-c) Websearch, (d) Datamining, (e) Educational and (f) Private DC workloads, respectively.}
	\label{fig:onenobackg}
	  \vspace{-1.5em}
\end{figure*}

In this section, we study the performance of T-RACKs implementation as a loadable Linux kernel module using synthetic workloads reproduced from the statistics of workloads found in production data-centers~\cite{Alizadeh2010, Greenberg2009}. T-RACKs is a shim-layer between the VMs (or TCP/IP stack) and the hypervisor (or link-layer). We use the NetFilter framework \cite{netfilter}, which is an integral part of Linux kernel. The NetFilter hooks attach to the data-path between the NIC driver and TCP/IP stack, which imposes no modifications to the TCP/IP stack of the host OS nor the guest OS. The module intercepts TCP packets incoming to the host or its guests before it is handed to the TCP/IP stack (i.e., at the post routing). First, the 4-tuples are hashed, and the associated flow index is calculated via Jenkins hash (JHash) \cite{jhash}. Then, TCP headers are examined, and the proper course of action is based on the flag bits (i.e.,  SYN-ACK, FIN, or ACK) following the logic in Algorithm~\ref{algo:rack1}. Unlike SNOOP~\cite{Balakrishnan1995}, the module does not employ any packet queues to store the incoming packets, it only stores and updates flow entry states (i.e., ACK No, arrival time and so on). Also, unlike \cite{Vasudevan2009} T-RACKs does not need the fine-grained high-resolution timers in the microsecond time-scale, therefore the native OS $Jiffies$ timer is used. T-RACKs uses a single timer for all flows to handle per-flow RTO events. These design choices make T-RACKs lightweight and help reduce the server overhead.

From 14 data-center grade servers equipped each with 6 NICs, we built a small-scale testbed consisting of 84 virtual servers, each assigned a dedicated physical NIC. The servers are interconnected via four non-blocking leaf switches and one spine switch. The testbed is organized into four racks (rack 1, 2, 3, and 4).  The servers are connected to leaf switches, and leaf switches are connected to the spine switch via 1 Gbps Ethernet links. The servers use Ubuntu Server 14.04 LTS with Linux kernel 3.18, which has integrated a full implementation of DCTCP. Unless otherwise stated, T-RACKs runs with the default settings (i.e., The RTO and threshold $\gamma$ of T-RACKs is set to 4 ms and 100 KB, respectively). The RTO of 4 ms is a reasonable 16 times (i.e., $\geq\!10$) the average RTT of $\approx\!250\mu s$ without queuing. We use our custom-built traffic generator to run the experiments with realistic traffic workloads. The traffic generator generates common workloads described in the literature (e.g., industrial-like Websearch~\cite{Alizadeh2010}, Datamining~\cite{Greenberg2009} or institutional-like University and Private DC~\cite{Benson2010-2}). In addition, we have deployed the iperf program \cite{iperf} to emulate large background traffic (e.g., VM migrations, backups) in some scenarios. We use different scenarios to reproduce one-to-all and all-to-all flows with or without background traffic. In the one-to-all scenarios, randomly chosen clients in one rack send random requests to any of all the servers in the data-center. While in the all-to-all scenario, all clients in the data-center send requests to randomly picked servers out of all the servers in the data-center. If background traffic is introduced, we run large iperf flows from all clients to all servers to evaluate T-RACKs under sudden and persistent network load spikes. Like before, we classify flows of size $\leq100KB$ as small,  [$100KB$-$10MB$] as medium, and $\geq10MB$ as large.

\begin{figure}[t]
\captionsetup[subfigure]{justification=centering}
\centering
	\begin{subfigure}[ht]{0.45\columnwidth}
		\includegraphics[width=\textwidth]{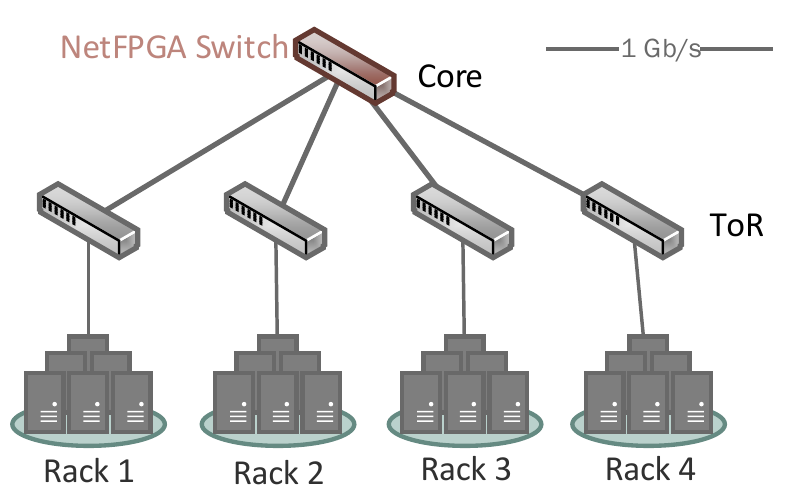}
		\caption{The testbed topology}
	\end{subfigure}
	\hfill
	\begin{subfigure}[ht]{0.45\columnwidth}
		\includegraphics[width=\textwidth]{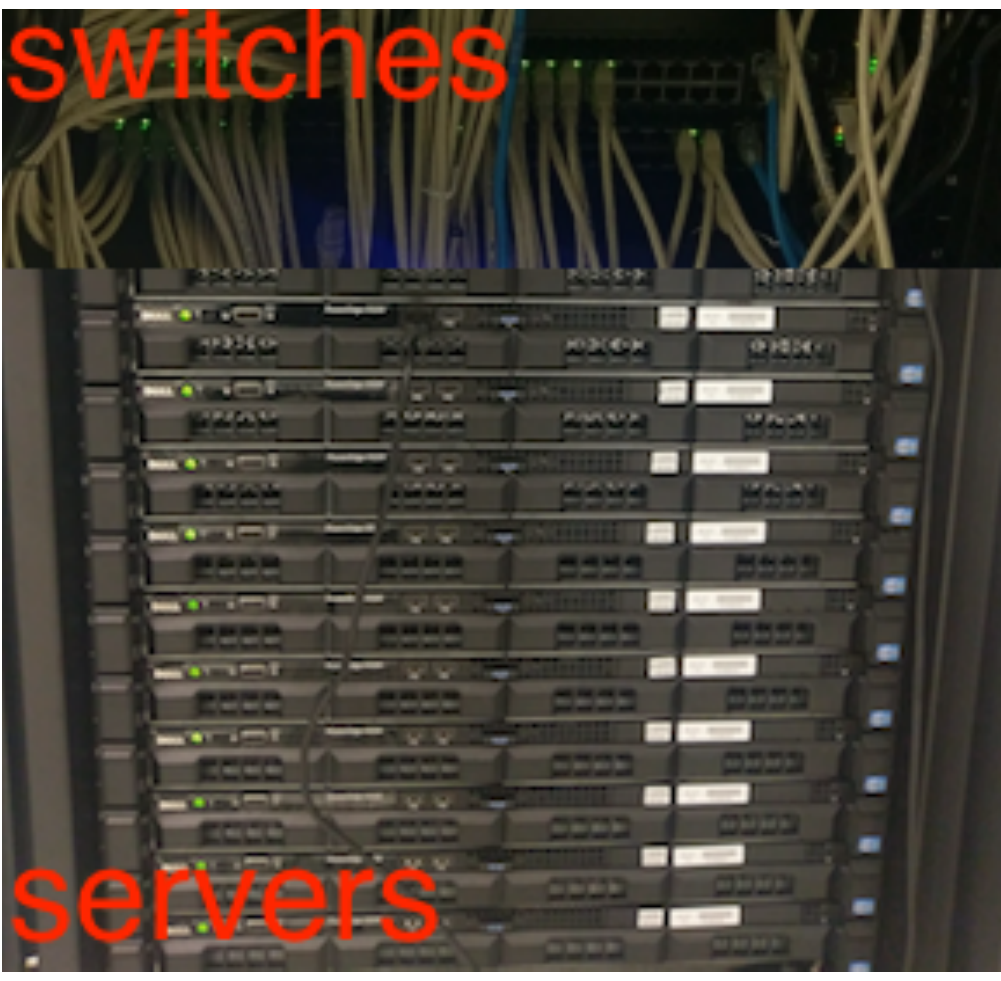}
		\caption{The actual testbed}
	\end{subfigure}
	\caption{Testbed setup of T-RACKs in small-scale cluster}
	\label{fig:RACKtestbed}
\end{figure}

\subsection{Experimental Results and Discussion}

\begin{figure*}[ht]
\captionsetup[subfigure]{justification=centering}
\centering
	\begin{subfigure}[ht]{0.32\textwidth}
             \includegraphics[width=\textwidth]{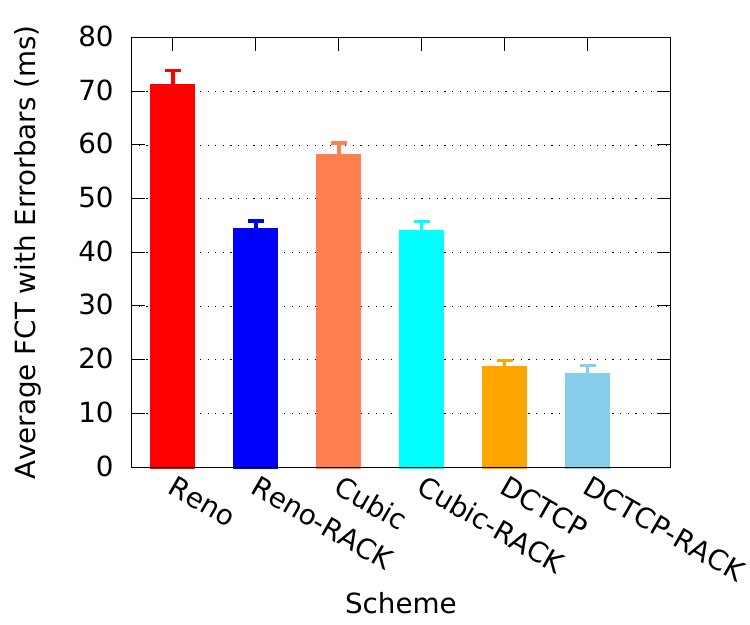}
      \caption{Small Flows: Average FCT}
              \label{fig:smallavgfct2}
       \end{subfigure}
       \hfill
       \begin{subfigure}[ht]{0.32\textwidth}
             \includegraphics[width=\textwidth]{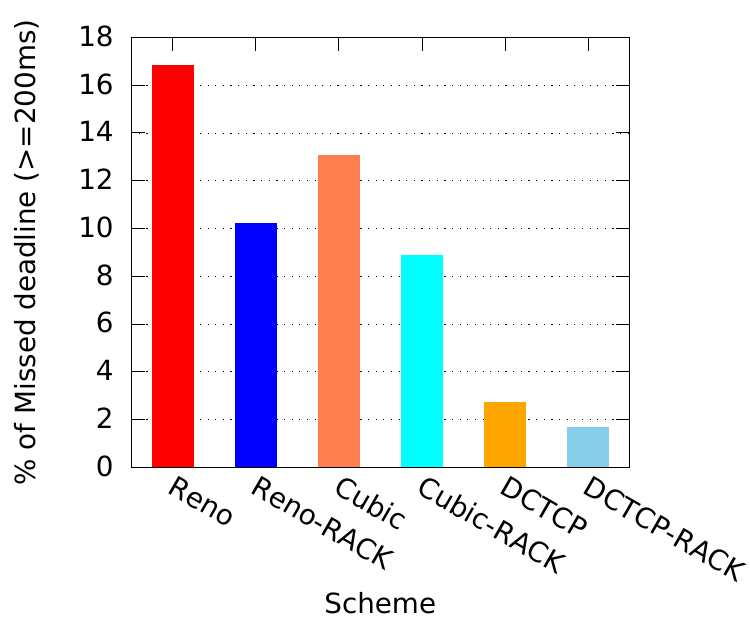}
      \caption{Small Flows: Missed Deadlines}
              \label{fig:smallmiss2}
       \end{subfigure}
	\hfill
	\begin{subfigure}[ht]{0.32\textwidth}
           \includegraphics[width=\textwidth]{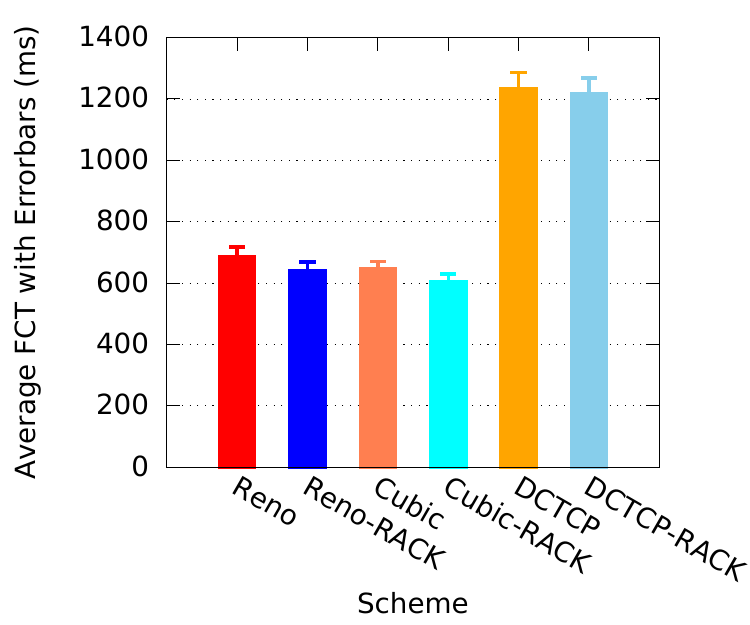}
	\caption{All Flows: Average FCT}
     \label{fig:allavgfct2}
       \end{subfigure}
		\\
       \begin{subfigure}[ht]{0.32\textwidth}
             \includegraphics[width=\textwidth]{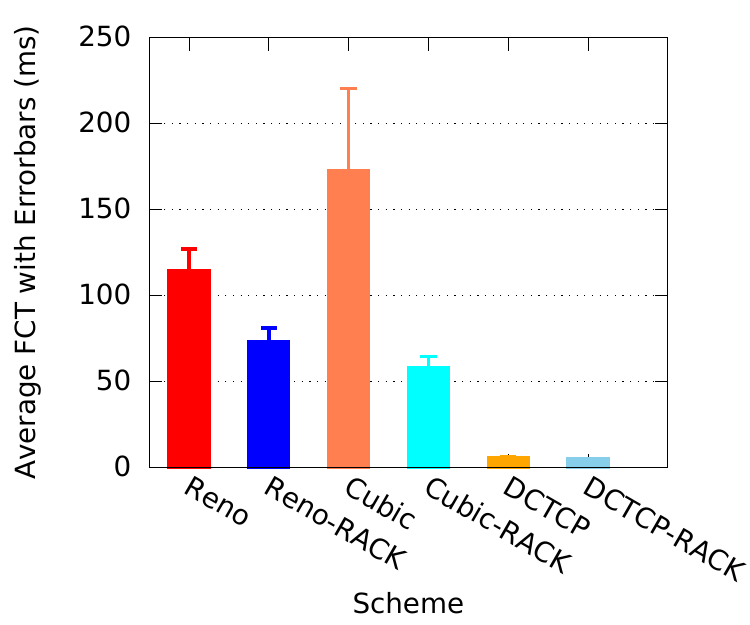}
      \caption{Average FCT (Datamining)}
              \label{fig:smallavgfctdatmin2}
       \end{subfigure}
       \hfill
       \begin{subfigure}[ht]{0.32\textwidth}
             \includegraphics[width=\textwidth]{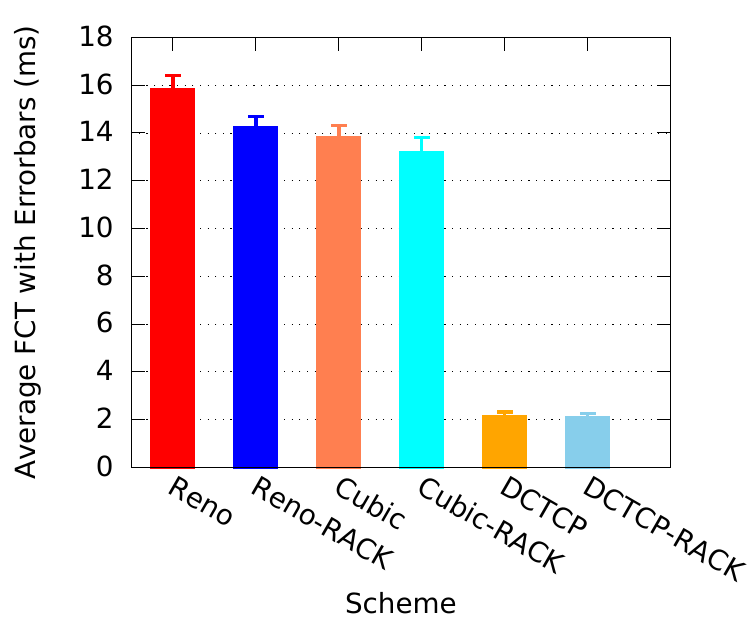}
             \caption{Average FCT (Educational)}
              \label{fig:smallavgfctedu2}
       \end{subfigure}
	\hfill
	\begin{subfigure}[ht]{0.32\textwidth}
           \includegraphics[width=\textwidth]{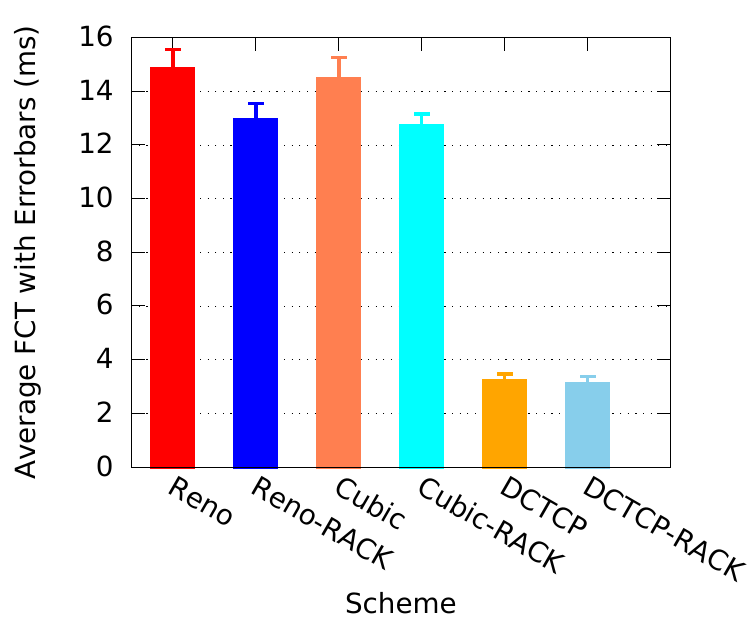}
	    \caption{Average FCT (Private DC)}
     \label{fig:smallavgfctprv2}
       \end{subfigure}

\caption{Performance metrics of one-to-all scenario with background traffic for (a-c) Websearch, (d) Datamining, (e) Educational and (f) Private DC workloads, respectively.}
	\label{fig:onebackg}
\end{figure*}

\textbf{One-to-all Scenario without Background Traffic: } we report here the average FCT for small and all flows as well as the number of small flows that missed a deadline of 200ms.%

\begin{figure}[t]
\captionsetup[subfigure]{justification=centering}
\centering
		 \begin{subfigure}[ht]{0.45\columnwidth}
        \includegraphics[width=\textwidth]{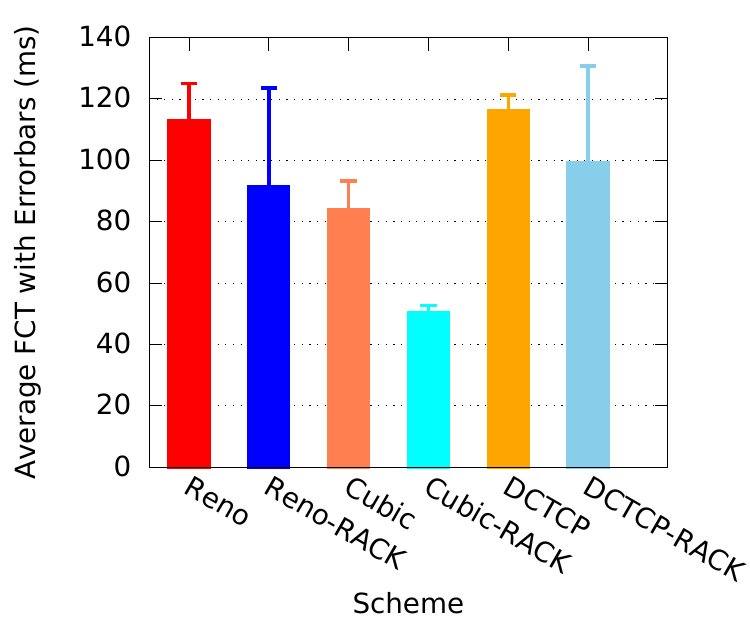}
      \caption{Websearch}
        \label{fig:smallavgfctweb3}
        \end{subfigure}
		\hfill
       \begin{subfigure}[ht]{0.45\columnwidth}
         \includegraphics[width=\textwidth]{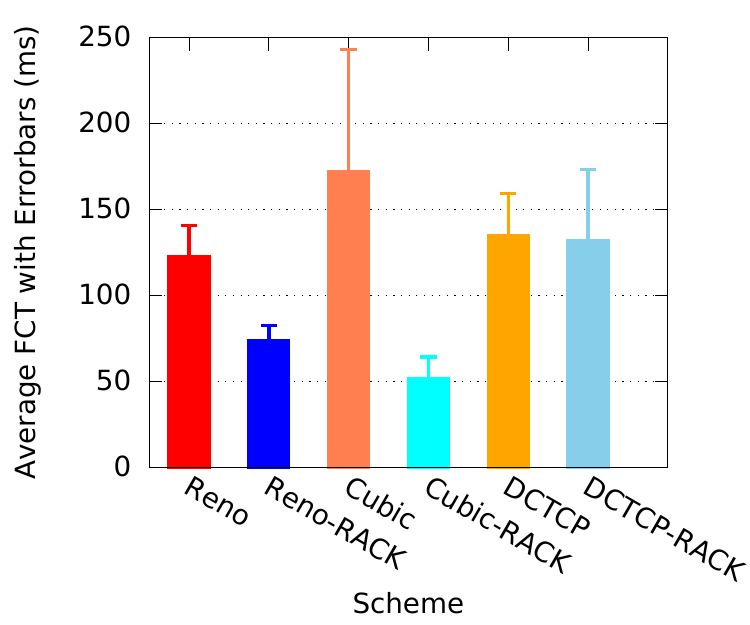}
      \caption{Datamining}
              \label{fig:smallavgfctdatmin3}
       \end{subfigure}
       \\
       \begin{subfigure}[ht]{0.45\columnwidth}
             \includegraphics[width=\textwidth]{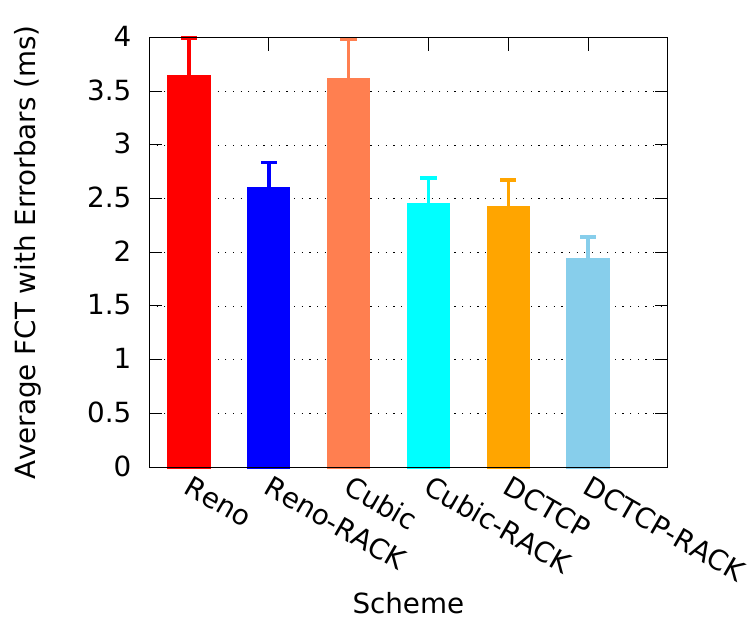}
             \caption{Educational}
              \label{fig:smallavgfctedu3}
       \end{subfigure}
	\hfill
	\begin{subfigure}[ht]{0.45\columnwidth}
     \includegraphics[width=\textwidth]{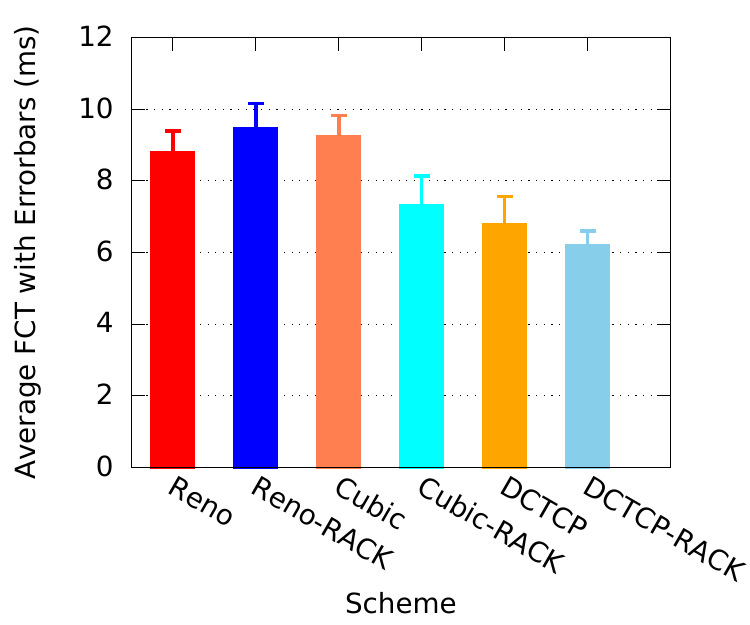}
	    \caption{Private DC}
     \label{fig:smallavgfctprv3}
       \end{subfigure}

\caption{The average FCT of small flows in the all-to-all scenario for (a) Websearch, (b) Datamining, (c) Educational and (d) Private DC workloads, respectively.}
	\label{fig:allnobackg}
\end{figure}

The traffic generator is deployed on every client running on an end-host in the data-center and is set to randomly initiates 1000 requests to randomly chosen servers on any of the other racks. In the Websearch workload, Figures~\ref{fig:smallavgfct1},~\ref{fig:smallmiss1}~and~\ref{fig:allavgfct1} show the average FCT and missed deadlines for small flows and the average FCT for all flows, respectively. While, Figures~\ref{fig:smallavgfctdatmin1},~\ref{fig:smallavgfctedu1},~and~\ref{fig:smallavgfctprv1}, show the average FCT for small flows in the Datamining, Educational, and private DC workloads, respectively. From these figures, we make the following observations:
\begin{inparaenum}[\itshape i) \upshape]
\item for all workloads, T-RACKs helps small flows regardless of the TCP version, on both the average FCT and its variation, as indicated by the error bars. Compared to Reno, Cubic and DCTCP, T-RACKs reduces the average FCT of small flows by $\approx(34\%, 49\%, 19\%)$ for Websearch, $\approx(18\%, 29\%, -)$ for Datamining, $\approx(69\%, -,35\%)$ for Educational and $\approx(-, - , 22\%)$ for Private DC workloads. We notice that DCTCP improves the FCT over its Reno and Cubic counterparts, and T-RACKs could improve their performance in terms of missed deadlines in Websearch. The average FCT, in certain cases of Educational and Private DC workloads, shows a negligible increase of FCT with T-RACKs. In these workloads, the network load is very light (as shown by the small FCT without T-RACKs), and hence the added overhead of deploying T-RACKs module surpasses its performance gains for these light workloads.
\item for Websearch workload, T-RACKs reduces the missed deadlines for short flows by $\approx(55\%, 53\%, 35\%)$ for Reno, Cubic, and DCTCP, respectively. 
\item T-RACKs slightly improves the overall average FCT. This can be attributed to the fact that small flows are finishing their transmission quicker, leaving some additional bandwidth for medium and large flows. The improvement shown equals $\approx(16\%, 5\%)$ for Reno and Cubic, respectively. In~Figure~\ref{fig:allavgfct1}, DCTCP with T-RACKs, shows a slight increase in average FCT of all flow types for Websearch workload which has many flows of medium size compared to other workload (Figure~\ref{fig:sizeinter}). While DCTCP is designed for improving FCT of small flows in DC environments, T-RACKs is designed to curb RTO events to improve the FCT of small flows whose transmitted volumes do not exceed the threshold $\gamma$. Hence, T-RACKs might be adding a slight overhead due to the need to maintain the flow table information for the medium and/or large flows. This overhead may be mitigated  by simply skipping state maintenance for flows that exceed the threshold and become long-lived. Moreover, the overhead for TCP variants, which are designed for Internet (e.g., Reno and Cubic), is nearly negligible relative to the large FCT of their long-lived flows. %
\end{inparaenum}\\
\textbf{One-to-all Scenario with Background Traffic: } to put T-RACKs under true stress, we run the same one-to-all scenario with all-to-all background traffic. Figure~\ref{fig:smallavgfct2}, Figure~\ref{fig:smallmiss2} and Figure~\ref{fig:allavgfct2} show the average FCT and missed deadlines for small flows as well as the average FCT for all flows for Websearch and Figure~\ref{fig:smallavgfctdatmin2}, Figure~\ref{fig:smallavgfctedu2} and Figure~\ref{fig:smallavgfctprv2} show the average FCT for short flows for data mining, educational, and private DC workloads, respectively. We observe the following: 
\begin{inparaenum}[\itshape i) \upshape]
\item T-RACKs can improve the average FCT of small flows for all workloads regardless of the TCP congestion control in use. As shown in the figures, compared to Reno, Cubic and DCTCP, T-RACKs reduces the average FCT of small flows by $\approx(38\%, 25\%, 7\%)$ for Websearch, $\approx(11\%, 5\%, 3\%)$ for educational and $\approx(13\%, 13\%, 4\%)$ for private DC workloads. The improvement increases for Datamining workload to $\approx(36\%, 67\%, 14\%)$ since it includes a wider range of short flows. 
\item T-RACKs reduces the missed deadlines for short flows of Websearch by $\approx(40\%, 33\%, 39\%)$ for Reno, Cubic, and DCTCP, respectively. 
\item T-RACKs still improves for the overall average FCT $\approx(7\%, 5\%, 2\%)$ for Reno and Cubic, and DCTCP respectively. 
\end{inparaenum}\\
\textbf{All-to-all Scenario without Background Traffic: }  we run the all-to-all scenario where all clients initiate 1000 requests each to any of all the servers in the data-center. Figure~\ref{fig:smallavgfctweb3}, Figure~\ref{fig:smallavgfctdatmin3}, Figure~\ref{fig:smallavgfctedu3} and Figure~\ref{fig:smallavgfctprv3} show the average FCT for short flows in Websearch, Datamining, Educational, Private workloads, respectively. The network load is considerably higher than the previous cases, given the more complex nature of this all-to-all traffic. We can still see here that T-RACKs can deliver significant improvements of up to 71\% in the FCT for all workloads. 

In summary, the experimental results show the performance gains achieved by T-RACKs, especially for small flows, that constitute the lion's share in data-centers traffic, without affecting too much the performance of larger flows. In particular, the results show that:
\begin{itemize}
\item T-RACKs reduces the variance of small flows' FCTs and the missed deadlines.
\item T-RACKs can maintain its gains even if bandwidth-greedy large flows hog the network.
\item T-RACKs efficiently handles various workloads, and is agnostic to the variant of TCP congestion controller.
\item T-RACKs fulfilled its requirements with no assumptions about nor any modifications to in-network hardware nor the TCP/IP stack of the guest VMs.
\end{itemize}

\section{Conclusion and future work}\label{sec:5}
\label{sec:conclude}

In this paper, we studied packet losses and the impact of various recovery methods on flow performance. We then proposed T-RACKs, an efficient cross-layer approach for timely recovery from losses. T-RACKs improves the flow completion time of time-sensitive flows and helps avoid throughput-collapse situations. T-RACKs is deployed either at the sender-side or the receiver-side as a shim-layer residing between the virtual machines and the network hardware. Simulation and experimental results show that the flows completion time is improved by up to an order of magnitude, missed deadlines are reduced considerably, and a high-link utilization is attained. T-RACKs is shown to be lightweight and practical due to its minimal footprint on end-hosts. Finally, because it does not change TCP and adapts to any TCP flavor, T-RACKs is very appropriate for multi-tenant public data-centers. As part of our future work, we seek real larger-scale deployment in cloud environments such AWS or Azure and investigate and analyze the effectiveness in T-RACKs scheme at scale.

\balance
\bibliographystyle{abbrv}
{
\bibliography{references}
}

\end{document}